\documentclass[11pt,a4paper]{article}
\usepackage{t1enc}
\usepackage[latin1]{inputenc}
\usepackage[english]{babel}
\usepackage[numbers]{natbib}
\usepackage{amssymb,stmaryrd,wasysym}
\usepackage{graphicx}
\usepackage[final]{pdfpages}
\usepackage{caption}

\newcommand{\beq}{\begin{equation}}
\newcommand{\eeq}{\end{equation}}
\newcommand{\beqarr}{\begin{eqnarray}}
\newcommand{\eeqarr}{\end{eqnarray}}
\newcommand{\beqa}{\begin{eqnarray*}}
\newcommand{\eeqa}{\end{eqnarray*}}
\begin{document}
\thispagestyle{empty}

\title{Clusters of galaxies in a Weyl geometric approach to gravity}
\author{Erhard Scholz\footnote{University of Wuppertal, Faculty  of Math. \& Nat.  Sciences and Interdisciplinary Centre for Hist. and Phil.  of Science; \quad  scholz@math.uni-wuppertal.de}}
\date{Nov 22, 2016 } 
\maketitle
\begin{abstract}
 A  model for the dark  halos of galaxy clusters,  based on  the    Weyl geometric scalar tensor theory of gravity (WST) with a MOND-like approximation, is proposed. It is uniquely determined by the baryonic mass distribution of hot gas and stars. A first heuristic check against  empirical data  for 19 clusters (2 of which  are  outliers), taken from the literature, shows encouraging results. Modulo a  caveat resulting from  different background theories (Einstein gravity plus $\Lambda CDM$ versus WST), the total mass for 14 of the   outlier reduced ensemble of 17 clusters seems to be predicted correctly (in the sense of overlapping $1\,\sigma$ error intervals).  \\
\end{abstract}

{\small \tableofcontents}

\section*{Introduction}
\addcontentsline{toc}{section}{\protect\numberline{}Introduction}
{\em This is a {\sc  corrected}, and in some lesser aspects improved, {\sc  postprint} version of the paper of the same title, published in {\sc Journal of Gravity} (JG) volume 2016, article ID 9706704. The most important corrections refer to equs. (28) and (47) of the publication in JG. An error in  JG (47) lead to a wrong relation between the acceleration due to the scale connection  ($a_{\varphi}$) and the acceleration arising from the  scalar field energy density ($a_{sf}$). The corrected form of JG (47) appears here as equ. (\ref{SF energy density}) and has consequences for the model  (equs. (\ref{total acceleration})ff.). It makes a new  run of the data  evaluation necessary, now based on the corrected dynamical equations. The new results are  given  in the updated tables 2, 3  and figs. 2 to 7.  The overall picture of the empirical test does not change, although now three rather than two galaxy clusters agree with the model only  in the $2\sigma$ range.}\\[0.5em]

In this paper  the gravitational dynamics of galaxy clusters is investigated from the point of view of  Weyl geometric scalar tensor theory of gravity (WST) with a non-quadratic kinematic Lagrange term  for the scalar field (3L) similar to the first relativistic MOND theory rAQUAL (``relativistic a-quadratic Lagrangian'') \citep{Scholz:2015MONDlike}. To make  the   paper as self-contained as possible, it starts with an outline of WST-3L  (section 1). WST-3L  has two (inhomogeneous) centrally symmetric  static weak field approximations:  
(i) the  Schwarzschild-de Sitter solution with its   Newtonian approximation, which  is valid if  the scalar field and the  WST-typical scale connection plays a negligible role; (ii)  a MOND-like approximation which  is appropriate under the constraints that  the scale connection cannot be ignored but is still small enough to allow for a Newtonian weak field limit of the (generalized) Einstein equation. The acceleration in the MOND approximation consists of a Newton term and an additional acceleration of which three quarters are due to  the energy density of the scalar field and one quarter  to the scale connection typical for Weyl geometric gravity. 

In centrally symmetric constellations the {\em scalar field} energy forms a {\em halo} about the baryonic mass concentrations. Besides the acceleration derived from the (Riemannian) Levi-Civita connection induced by the baryonic matter and the scalar field energy an {\em additional acceleration} component  due to the Weyl geometric scale connection arises in the present approach.  If the latter is expressed by a fictitious mass in Newtonian terms, a {\em phantom halo} can be ascribed to it. It indicates the amount of mass one has to assume in the framework of Newton dynamics to produce the the same amount of additional acceleration. The scalar field halo consists of true energy derived from the energy-momentum tensor; it is  independent of the reference system, as long as  one restricts the consideration to reference systems with low (non-relativistic) relative velocities. The phantom halo, on the other hand,  is a symbolical construct and valid only in the chosen reference system (and scale gauge).  
For galaxy clusters we  find   two components of the scalar field halo, one deriving from  the total baryonic mass in the MOND approximation of the barycentric rest system of the cluster (component 1), and one arising from the superposition of all the scalar field halos forming around each single galaxy in the MOND approximation of the latter's rest system (component 2).
Because velocities of the galaxies with regard to  the cluster barycenter are small (non-relativistic) and also the energy densities are small, the two components can be superimposed additively (linear approximation). 
With regard  to the barycentric rest system of the cluster   a {\em three component halo for clusters of galaxies} arises, two components being
 due to the scalar field energy and one  purely phantom  (section 2).

The two component scalar field halo is a distinctive feature of the Weyl geometric scalar tensor approach; it is neither present in the non-relativistic MOND  approaches nor in rAQUAL. One may pose the question whether it suffices for explaining the deviation of   the cluster dynamics from the Newtonian expectation without    additional dark matter.  
If we call the totality of the  three components the {\em transparent halo} of the cluster (section 2.5), the question is whether the (theoretically derived) transparent halo can explain the {\em dark halo} of galaxy clusters, observationally determined in the framework of Einstein gravity and $\Lambda CDM$.

Section 3 contains a first test of the model by confronting it with empirical data on mass distribution available in the astronomical literature. A full-fledged test would presuppose an evaluation  of raw observational data in the framework of the present approach and is beyond the scope of this study. Here we use  recently published data on the total mass (dark plus baryonic), hot gas, and the star matter  for 19 galaxy clusters, which have been determined from different observational data sources (and are thus more precise than earlier ones) \citep{Reiprich/Zhang_ea:2011}, \citep{Lagana/Reiprich_ea:2011}, \citep{Reiprich/Zhang_ea:Corr}.
    2 of the 19 clusters show a surprisingly large relation of total mass to gas mass. They are separated as outliers from the rest of the ensemble already by the authors of the study; so do we. 17 non-outlying clusters remain as our core reference ensemble. 

 In the  mentioned studies total  mass, gas mass and star mass are  determined on the background of Einstein gravity plus $\Lambda CDM$. That raises the  problem of  compatibility with the WST framework. It is discussed in  3.1,  3.2 and leads to a certain caveat with regard to the empirical values for the total mass ($M_{200}, \, M_{500}$) and the reference distances $r_{200}, \, r_{500}$ to the cluster centers. But it does not seem to obstruct the possibility for a first empirical  check of our model (section \ref{subsection theory dependence}). More refined studies are welcome. They have to use the WST framework for evaluating the observational raw data or, at least, to analyze the transfer problem of mass data from one framework to the other in more  detail.

For 14 of the 17 main reference clusters the empirical and the theoretical values for the total  mass agree in the sense of overlapping $1\,\sigma$ error intervals. For the remaining three we find overlap  in the  $2\,\sigma$ range. The 
two  outliers of the original study do not lead to overlapping intervals   even in the  $4\,\sigma$ range (section \ref{subsection 19 clusters}).  In the present approach the dynamics of the {\em Coma cluster is explained  without assuming} a component of {\em particle dark matter}. It is being discussed in more detail than the other clusters  in section \ref{subsection Coma}.

 A short comparison with the halos of R. Sanders'  $\mu_2$-MOND model with an additional neutrino core \citep{Sanders:2003}, and with the NFW halo \citep{Geller_ea:1999} is given for Coma (section \ref{subsection comparisons}).
The paper is rounded off by  a short remark on the bullet cluster
 (section \ref{subsection bullet}) and a final discussion (section \ref{section discussion}).

\section{Theoretical framework   \label{section framework}}

\subsection{\small Weyl geometric scalar tensor theory  of gravity  (WST)\label{subsection WST}}
Among the family of scalar tensor theories of gravity the best known ones, and closest to Einstein gravity, are those with a Langrangian containing a  modified Hilbert term coupled to a scalar field $\phi$. Their Lagrangian has the general form
\beqarr L &=& \frac{1}{2} (\xi \phi)^2 R + L_{\phi} -  \frac{\lambda}{4} \phi^4\ldots \label{modified Hilbert term}\\
   \mathfrak{L} &=& L \sqrt{|g|}\, , \qquad |g|=   |det\, g|\, . \nonumber
\eeqarr 
Here $g$ is an abbreviation for a 4-dimensional pseudo-Riemannian metric $g=(g_{\mu \nu})$ of signature $(-+++)$. $\phi$ is a real valued scalar field on spacetime, $ L_{\phi}$ its kinetic term, $\xi$ a constant allowing for bridging the  hierarchy between the Planck scale and cosmologically small quantities (like vacuum energy density), and the dots  indicate   matter and  interaction terms.
Under conformal rescaling of the metric, 
\beq g_{\mu \nu} \mapsto g'_{\mu \nu} = \Omega^2 \, g_{\mu \nu}  \quad \mbox{ ($\Omega$ a positive real valued function),  } \label{conformal rescaling}
\eeq 
  the scalar field changes with weight $-1$, i.e. $\phi \mapsto \phi'=  \Omega^{-1}\, \phi$. So far this is similar to the well known 
Jordan-Brans-Dicke (JBD)  scalar-tensor theory of gravity {\citep{Jordan:Schwerkraft}, \citep{Brans/Dicke}, \citep{Fujii/Maeda},  \citep{Faraoni:Frames}. 
But here we work in a scalar-tensor theory in the framework of Weyl's generalization of Riemannian geometry \citep{AGS,Poulis/Salim:2011,%
Quiros:2014,Scholz:2014paving,Scholz:2015MONDlike}. 

Crucial  for the {\em Weyl geometric scalar tensor  approach} (WST)  is  that 
 the scalar curvature $R$ and all  dynamical terms involving covariant derivatives are  expressed in Weyl geometric { scale covariant} form.\footnote{Fields $X$ are scale covariant if they transform under rescaling by $X \mapsto \tilde{X}= \Omega^w X$, with $w\in Q$ (in most cases even in $Z$); $w$ is called the weight of $X$. Covariant derivatives of scale covariant fields are defined such that the result of covariant derivation $D_{\mu}X$  is again scale covariant of the same weight $w$ as $X$.}
The  Lagrangian density $\mathcal{L}$ is  {\em invariant under conformal rescaling} for any value of the coefficient $\xi^2$ of the modified Hilbert term. 
For the matter and interaction terms of the standard model of elementary particles,  scale invariance is naturally ensured by the  coupling to the Higgs field which has the same rescaling behaviour as the gravitational scalar field. Although there is no complete identity, there is a close relationship between scale  and conformal invariance in quantum field theory \citep{Nakayama:2014}.
For classical matter we expect that a better understanding of the quantum to classical transition, e.g. by the decoherence approach, allows to consider scale invariant   Lagrangian densities also. For the time being we introduce the scale invariance of matter terms in the Lagrangian as a {\em postulate}. 

In our context an important consequence is the scale covariance of the Hilbert energy momentum tensor 
\beq T_{\mu \nu}=-  \frac{2}{ \sqrt{|g|}}\frac{\delta \mathcal{L}_{m}}{\delta g^{\mu\nu}}\, , \label{matter energy tensor}
\eeq 
which is of  weight $w(T_{\mu \nu})=-2$. That  is consistent with dimensional considerations on a phenomenological level.
It has been shown that the matter Lagrangian of quantum matter (Dirac field, Klein Gordon field) is consistent with test particle motion along geodesics (autoparallels) $\gamma(\tau)$ of the affine connection,  if the underlying Weyl geometry is integrable (see below, equ. (\ref{varphi})) \citep{AGS}. For classical matter we assume the same. It is to be expected that  it can be proven similar to Einstein gravity.

We do not want to heap up too many technical details; more can be found in the literature given above. But we  have to mention that  a {\em Weylian metric} can be given by an equivalence class of pairs $(g, \varphi)$ consisting of  a pseudo-Riemannian metric $g=g_{\mu \nu}dx^{\mu} dx^{\nu}$, the {\em Riemannian component} of the Weyl metric, and a differentiable one-form $\varphi = \varphi_{\mu}dx^{\mu}$,  the {\em scale connection} (or, in Weyl's original terminology, the ``length connection'').  $ \varphi_{\mu}$ is often called the Weyl covector or even Weyl the vector field.
 In fact, $\varphi$ denotes a connection with values in the Lie algebra of the scale group $(R^+, \cdot )$ and can locally be represented by  a differentiable 1-form. 
The equivalence is given by rescaling  the Riemannian component of the Weylian metric according to (\ref{conformal rescaling}), while 
 $\varphi$ has the peculiar gauge transformation behaviour of a {\em connection}, rather than that of an ordinary vector (or covector) field in a representation space of the scale group:
\beq \varphi_{\mu} \mapsto \varphi'_{\mu} =  \varphi_{\mu} - \frac{\partial_{\mu} \Omega}{\Omega}\, \label{gauge transformation}
\eeq
or, shorter, $ \varphi' =  \varphi - d \log \Omega $.

A Weylian metric has a uniquely determined compatible affine connection $\Gamma$; in physical terms it characterizes the {\em  inertio-gravitational} guiding { field}. It can be additively composed by  the well known Levi-Civita connection $_g\hspace{-0.12em}\Gamma $ of the Riemannian component $g$ of any gauge $(g, \varphi)$  and an additional expression  $_{\varphi}\hspace{-0.15em}\Gamma $ in the scale connection, in short
\beq \Gamma = \, _g\hspace{-0.12em}\Gamma + _{\varphi}\hspace{-0.2em}\Gamma \quad \mbox{with} \quad  {}_{\varphi}\Gamma^\mu _{\nu \lambda }=    \delta ^{\mu }_{\nu } \varphi _{\lambda } +
\delta ^{\mu }_{\lambda } \varphi _{\nu } - g_{\nu \lambda } \varphi^{\mu }.\label{Levi-Civita}  
\eeq

Covariant derivatives $D$ in the Lagrangians (\ref{modified Hilbert term}),  (below)  (\ref{cubic Lagrangian}), and consequently in the expression for the energy-momentum tensor of the scalar field (\ref{ThetaI}, \ref{ThetaII}) below, denote those of Weyl geometry. For a covariant field  $X^{\nu}$ of weight $w$ the  derivativation $\nabla$ with regard to $\Gamma$  of (\ref{Levi-Civita}) $D$ is supplemented by a term due to the scaling weight $w$ of $X$:
\beq D_{\mu} X^{\nu} = \nabla_{\mu}X^{\nu} + w\, \varphi_{\mu}X^{\nu}= \partial_{\mu} X^{\nu} + \Gamma_{\mu \lambda}^{\nu}X^{\lambda} + w\, \varphi_{\mu}X^{\nu}. \label{covariant derivative}
\eeq 
It turns out that for the metric $g_{\mu \nu}$ the full covariant derivative is zero:
\beq D_{\lambda}g_{\mu \nu} = 0 \quad \longleftrightarrow \nabla_{\lambda}g_{\mu \nu} + 2 \varphi_{\lambda}g_{\mu \nu}=0\, 
\eeq
This is the Weyl geometric compatibility condition between metric and affine connection (sometimes called ``semi-metricity'').

In the low energy regime there are physical reasons to constrain the scale connection to the {\em integrable case} with a closed differentiable form $d \varphi=0$, i.e. $\partial_{\mu}\varphi_{\nu}=\partial_{\nu}\varphi_{\mu}$. 
 Then $\varphi$ is  a gradient (at least locally) and may be given by 
\beq \varphi_{\mu}= - \partial_{\mu} \omega \, . \label{varphi}
\eeq 
This constraint is part of the defining properties  of WST.
Then it is  possible to ``integrate the scale connection away''  \citep{AGS}. Having done so,  the Weylian metric, given as $(\tilde{g},\tilde{\varphi})$, is characterized by its Riemannian component $\tilde{g}_{\mu \nu}$ only (and a vanishing scale connection $\tilde{\varphi}_{\mu}=0$). By obvious reasons we call this the {\em Riemann gauge}. This  is the analogue of the choice of Jordan frame in JBD theory.

In any case, the choice of a representative $g_{\mu \nu}$ also fixes  $\varphi_{\mu}$; both together define a {\em scale gauge} of the Weylian metric. Conformal rescaling of the metric is accompanied by the gauge transformation of the scale connection (\ref{gauge transformation}). From a mathematical point of view all the scale gauges are on an equal footing, and the physical content of a WST model can be extracted, in principle, from any scale gauge. One only needs to form a proportion with the appropriate power of the scalar field.
 From the physical point of view there are, however,  two particularly  outstanding scale gauges. Of special importance besides Riemann gauge is the  gauge in which the scalar field is scaled to a constant  $\phi_o$ ({\em scalar field gauge}). For the particular choice of the constant value such that
\beq
(\xi \phi_o)^2 = (8 \pi \, G)^{-1} = E_{pl}^2 \, , \label{Einstein gauge}
\eeq
 with the Newton gravitational constant $G$, this gauge is called {\em Einstein gauge} ($E_{pl}$ the reduced Planck energy). It is the analogue of Einstein frame in JBD theory. 
In this gauge the metrical quantities (scalar, vector or tensor components) of physical fields are  directly expressed by the corresponding field or field component of the mathematical model (without the necessity of forming  proportions). 

 In Riemann gauge    $\Gamma$ reduces to $ _g\hspace{-0.12em}\Gamma$ by definition. Thus, in this gauge, the  guiding field is given by the ordinary expression of the Levi-Civita connection.  On the other hand, in Einstein gauge the measuring behaviour of clocks are most immediately represented by the metric field; and also other physical observables are most directly expressed by the field values in this scale. Then the expression of the  gravitational field and with it the expressions for accelerations contain contributions from the Weylian scale connection. Thus a specific dynamical difference to Einstein gravity and Riemannian geometry (as well as to JBD theory) arises even in the case of  WST with its integrable Weyl geometry.

Writing the scalar field $\tilde{\phi}$ in Riemann gauge $(\tilde{g},0)$ in exponential form,
$\tilde{\phi} = e^{\omega}$,  turns its exponent
\beq
\omega:= \ln{\tilde{\phi}} \label{tilde omega}
\eeq
into  a {\em scale invariant} expression for the scalar field. In the following  we shall omit the tilde sign to simplify notation.  The scale connection $\varphi = \hat{\varphi}$ in scalar field gauge  is then
\beq \hat{\varphi} = - d  \omega \, , \label{eq check varphi}
\eeq 
because $\Omega=\tilde{\phi}=  e^{\omega}$ is the rescaling function from Riemann gauge to scalar field gauge.
For more details see \citep{Adler/Bazin/Schiffer,Blagojevic:Gravitation,Poulis/Salim:2011,Quiros:2014,Scholz:2015MONDlike}.

 For the sake of consistency under rescaling we  consider {\em scale covariant geodesics} $\gamma(\tau)$ with scale gauge dependent parametrizations of the geodesic curves of weight $w(\dot{\gamma})=-1$: 
\beq \dot{u}^{\lambda} + \Gamma_{\mu \nu }^{\lambda}u^{\mu}u^{\nu} - \varphi_{\mu}u^{\mu}u^{\lambda} = 0 \, , \qquad u^{\mu}= \dot{\gamma}^{\mu}\,   \label{eq covariant geodesic}
 \eeq 
Here the affine connection contains  a 
 $\varphi$-dependent term in addition to the well-known  Levi-Civita connection $_{\varphi}\hspace{-0.1em}\Gamma^{\lambda}_{\mu \nu }$ derived from  $g_{\mu\nu}$ (equ (\ref{Levi-Civita})). The last term on the l.h.s. of (\ref{eq covariant geodesic}) takes care for the scale dependent parametrization (compare (\ref{covariant derivative})). In this way we work with a {\em projective} family of paths.

For any gauge of the Weylian metric and the scalar field, $(g, \varphi,\phi)$,   any timelike geodesic  has thus a generalized proper time parametrization $\gamma(\tau)$ with $g_{\mu \nu}u^{\mu}u^{\nu}=-1$, where $u^{\mu}=\dot{\gamma}^{\mu}$. 
Inverting  the coordinate time function $t(\tau )$ along the geodesic by $\tau(t)$ we have, in abbreviated notation, $\tau ' t' = 1$ and thus:
\beq \frac{d^2 x^{i  }}{dt ^2} =
 \frac{d\tau}{dt } \frac{d}{d\tau} (\frac{d\tau}{dt } \frac{dx^{i  }}{d\tau}) = 
\left( \frac{d\tau}{dt }\right)^2 \frac{d^2 x^{i  }}{d\tau^2} -
 \left(\frac{dt }{d\tau}\right)^{-3}\frac{d^2 t }{d\tau^2}\frac{dx^{i  }}{d\tau} \, 
\eeq
With (equ. (\ref{eq covariant geodesic})) and indices $i,j,k=1,2,3$, $\mu, \nu \ldots=0,1,2,3$ this leads to 
\beqarr  \frac{d^2 x^{i  }}{dt ^2} &= & 
\left( \frac{d\tau}{dt }\right)^2  \left(  - \Gamma^{i  }_{\mu \nu } \frac{dx^\mu}{d\tau}\frac{dx^\nu }{d\tau} + \varphi_\mu \frac{dx^{i  }}{d\tau} \frac{dx^\mu}{d\tau}  \right) \nonumber \\
& & - \left( \frac{d\tau}{dt }\right)^3 \frac{dx^{i }}{d\tau} \left( -\Gamma^0_{\mu \nu } \frac{dx^\mu}{d\tau}\frac{dx^\nu }{d\tau} + \varphi_\mu \frac{dt }{d\tau} \frac{dx^\mu}{d\tau}  \right) \\
&=&    - \Gamma^{i  }_{\mu \nu } \frac{dx^\mu }{dt }\frac{dx^\nu }{dt } + \varphi_\mu  \frac{dx^{i  }}{dt } \frac{dx^\mu }{dt } 
+ \Gamma^0_{\mu \nu } \frac{dx^\mu }{dt }\frac{dx^\nu }{dt } \frac{dx^{i }}{dt } - \varphi_\mu   \frac{dx^\mu }{dt } \frac{dx^{i }}{dt } \, . \nonumber
\eeqarr
 \noindent
Happily, the length connection terms coming from the scale covariance modification of  the geodesic equation (\ref{eq covariant geodesic})  cancel. This is an expression of  the fact that only the trace of the geodesic enters into the coordinate time parametrization of the dynamical equation (\ref{equ of motion}).
The equation of motion for mass points in Weyl geometric gravity, parametrized  in coordinate time, becomes:
\beqarr \label{equ of motion}
\frac{d^2 x^{i  }}{dt^2} &= & 
- \Gamma^{i }_{00} +\Gamma^{0}_{00} \frac{dx^{i }}{dt}
 - 2 \Gamma^{i }_{0 j } \frac{dx^{j }}{dt}
-  \Gamma^{i }_{ j  k } \frac{dx^{j }}{dt}\frac{dx^{k }}{dt}  \\
& & + 2  \Gamma^{0}_{0 j  } \frac{dx^{i }}{dt}\frac{dx^{j }}{dt}
+ \Gamma^{0}_{ j  k  } \frac{dx^{i }}{dt}\frac{dx^{j }}{dt}\frac{dx^{k }}{dt}
\nonumber 
\eeqarr
In the result  the dynamics of mass points in Weyl geometric gravity is governed by  the guiding field (the affine connection), like in the  semi-Riemannian case 
 \citep[equ. (9.1.2)]{Weinberg:Cosmology}. Note, however, that in (\ref{equ of motion}) the {\em  length connection enters into the affine connection} and influences the dynamics because of (\ref{Levi-Civita}). 


 The geodesic equation thus  contains terms in the scale connection $\varphi_{\mu}$. In the low velocity, weak field regime the equation of motion reduces to the form well known from Einstein gravity $\frac{d^2 x^{j }}{dt^2} =  
- \Gamma^{j}_{00}$. Here the  $\Gamma^{j}_{00} =  \, _g\hspace{-0.12em}\Gamma^{j}_{00} + _{\varphi}\hspace{-0.35em}\Gamma^{j}_{00} $ ($j=1,2,3$) are the coefficients of the  Weyl geometric affine connection with $_{\varphi}\hspace{-0.1em}\Gamma^{j}_{00} $ given by (\ref{Levi-Civita}). This is the crucial  modifying term for  gravity in the Weyl geometric approach (low velocity case).


\subsection{\small   The weak field static approximation  \label{subsection weak field approximation}}

We want to understand the scale connection for the motion of point particles. The free fall of test particles  in Weyl geometric gravity follows scale covariant geodesics. It is governed by a differential equation  formally identical to the one in Einstein gravity  (\ref{equ of motion}). Here we look at the weak field static case for low velocities in order to study  the dynamics of stars in galaxies and galaxies in clusters. For studies of the gas dynamics and its modification in our framework the velocity dependent terms of (\ref{equ of motion}) have to be taken into account. This is not being done here.

Analogous to  Einstein gravity, the coordinate acceleration $a$  for a low velocity motion $x(\tau)$ in proper time parametrization is given  by
\beq a^{j}= \frac{d^2x^{j}}{d\tau^2} \approx - \Gamma^{j}_{oo}\, . \label{acc 1}
\eeq 
According to (\ref{Levi-Civita}) the total acceleration decomposes into 
\beq a^{j} = - _g\hspace{-0.1em}\Gamma^{j}_{oo} - _\varphi\hspace{-0.2em}\Gamma^{j}_{ \nu \lambda} =a^j_R + a^j_{\varphi} \,    \label{acceleration}
\eeq 
($j=1, 2, 3$  indices of the spacelike coordinates), 
where 
\beq a^j_R=- _g\hspace{-0.1em}\Gamma^{j}_{oo} \label{a_R}
\eeq
 is the Riemannian  component of the acceleration known from Einstein gravity. Clearly   
\beq
a^j_{\varphi}= - _\varphi\hspace{-0.15em}\Gamma^{j}_{oo} \label{add acc}
\eeq
  represents  an {\em additional acceleration}   due to the   Weylian scale connection.  For a diagonal Riemannian  metric  $g= diag\,(g_{oo}, \ldots, g_{33})$  the general expression  (\ref{Levi-Civita}) simplifies to $- _\varphi\hspace{-0.15em}\Gamma^{j}_{oo}= - g_{oo} \varphi^j$.
General considerations on observable quantities  and consistency with Einstein gravity show that, in order to confront it with empirically measurable quantities, we have to take its expression in {\em Einstein gauge} if we want to avoid additional rescaling calculations \citep[sec. 4.6]{Scholz:2014paving}.

For a (diagonalized) weak field approximation in Einstein gauge, 
 \beq 
 g_{\mu \nu} = \eta_{\mu \nu} + h_{\mu \nu}, \qquad |h_{\mu \nu}| \ll 1\, , \label{weak field}
 \eeq 
 with $\eta = \epsilon_{sig} \, \mbox{diag}(-1,+1,+1,+1)$, the Riemannian component of the acceleration is  the same as in Einstein gravity. Its leading term (neglecting 2-nd order terms in $h$) is,
 \beq
a^j_R =  \hspace{-0.1em}_g\Gamma^{j}_{oo} \; \approx \;  \frac{1}{2} \eta ^{jj}\partial_j h_{oo} \quad \mbox{(no summation over $j$)} \, . 
 \eeq
In the limit, $\Phi_N := -\frac{1}{2}h_{oo}$ behaves  like a Newtonian potential
\beq a_R \approx - \nabla \Phi_N \, ,  \label{Newton limit}
\eeq
where  $\nabla$ is understood to operate in the 3 spacelike coordinate space with  Euclidean coefficients as the leading term of the metric.  

 In Einstein gauge the Weylian scale connection $\hat{\varphi}_{\mu}$ arises  from Riemann gauge by rescaling with $\Omega= e^{\omega}$,   $\hat{\varphi}_{\mu}=-\partial_{\mu}\omega$ (\ref{gauge transformation}), and $ a^j_{\varphi} \approx \varphi^j$. In other words, the additional acceleration due to the scale connection (\ref{add acc}) is   generated by  the { scale invariant} representative $\omega$  of the {\em scalar field} as its {\em potential}:
\beq a_{\varphi}\approx - \nabla \omega \, 
\eeq 
If we compare with Newton gravity, we can calculate the fictitious mass density which one had to  assume  on the right hand side of the Poisson equation, in addition to the real masses, in order to generate the same amount of additional acceleration. Obviously here it is
\beq \rho_{ph} = (4 \pi G)^{-1}\, \nabla^2 \omega \, .
\label{phantom energy of scale connection}
\eeq 
 In the terminology of the MOND literature the acceleration due to the Weylian scale connection corresponds to a {\em phantom energy} density $\rho_{ph}$.\footnote{See, e.g., \citep[48]{Famaey/McGaugh:MOND}.} We see that already on the general level the dynamics of WST differs from Einstein gravity. Only for trivial scalar field, $\omega=const$, the usual {\em  Newton limit} is recovered, otherwise it is modified.  We shall explore how this modification  relates to the usual MOND approaches. 


\subsection{\small  WST gravity with cubic kinematic Lagrangian (WST-3L) \label{subsection WST-3L}}
The most common form of the kinetic term for the scalar field is that of a Klein-Gordon field, $L_{\phi 2}= -\frac{\alpha}{2} D_{\nu}\phi D^{\nu}\phi$, quadratic in the norm of the (scale covariant) gradient.\footnote{In WST  $D_{\nu}$ denotes the a scale covariant derivative of $\phi$, $D_{\nu}\phi = \partial _{\nu} \phi - \phi \, \varphi_{\nu}$.} 
For our form of the gravitational Lagrangian (\ref{modified Hilbert term}) it is conformally coupled for $\alpha=-6 \xi^2$.
Inspired by the  approach  of  the relativistic ``a-quadratic Lagrangian'' (rAQUAL),   the first relativistic attempt of a MOND theory of gravity \citep{Bekenstein/Milgrom:1984,Bekenstein:2004}, we find  
 that a Weyl geometric scalar tensor theory of gravity leads to a MOND-like phenomenology if we add a {\em cubic} term to the kinetic  Lagrangian of the scalar field
\beq L_{\phi} = L_{\phi 2} + L_{\phi 3} \, .
 \eeq
 The crucial difference to the early approach of rAQUAL is the  scale covariant  reformulation in the framework of Weyl geometry. It results in a different behaviour of the scalar field energy density.
Bekenstein/Milgrom's model relied crucially on implementing a  transition function $f(y)$ between the Newton and the deep MOND regime into the kinetic term. The constraint of scale invariance of the  Lagrangian reduces the underdetermination of the Lagrangian  and suggests a slightly different form of the kinetic term. It is still quite near to  the one of the relativistic AQUAL theory. 

 In the review paper \citep{Bekenstein:2004} Bekenstein gives the rAQUAL Lagrangian in the form (his eq. (6))
\beq L_{\psi} = - (8 \pi G_N)^{-1} L^{-2} f(L^2\, \partial_{\nu}\psi \, \partial^{\nu}\psi)\, , \label{Bekenstein's (6)}
\eeq
where $L$ is ``a constant with  dimension of length introduced for dimensional consistency'' (later it is identified as the MOND acceleration $a_o$ via $c^2\, L^{-1}= a_o$). Asymptotically $f(y) \sim \frac{2}{3}y^{\frac{3}{2}}$  for $y \ll 1$ ({\em MOND regime}); similarly  $f(y) \sim y$ for $y \gg 1$ ({\em Newton regime}).
 $\psi$ is the logarithm of a rescaling function between the Jordan frame metric $\tilde{g}_{\mu \nu}$ (called the ``physical'' metric) and the Einstein frame (``primitive'') metric $g_{\mu \nu}$, $\tilde{g}_{\mu \nu}=e^{2\psi}g_{\mu \nu}$. Its role is very close to our $\omega$ in (\ref{tilde omega}). A corresponding scale covariant form of the Lagrangian could be
\beq L_{\phi 3} =  (\xi \phi)^2 (\xi^{-1}\phi)^{2}\, f((\xi^{-1}\phi)^{-2} \, \partial_{\nu}\omega \,  \partial^{\nu}\omega)\, , \label{scale covariant rAQUAL}
\eeq
where  in Einstein gauge $\xi^{-1}\phi_o $ plays the role  of $a_o$ (up to a factor). The sign has deliberately been changed; the reasons are given below eq. (\ref{Milgrom equation}). 

We are here  interested in additive modifications (\ref{add acc}) of Einstein gravity, mainly in a domain in which the  effects of the scale connection $\varphi_{\nu}= -\partial_{\nu}\omega$, compared with  those of the Riemannian component of the metric,   cannot be neglected. This will be called a regime with {\em MOND approximation}. For  non-timelike $\nabla \omega$  the Lagrangian (\ref{scale covariant rAQUAL})  becomes, 
\beq L_{\phi 3} =  \frac{8}{3} \xi^3   \phi^{-2} \, (D_{\nu}\phi\, D^{\nu}\phi)^{ \frac{3}{2}}=   (\xi \phi)^2 ( \frac{1}{4}\xi^{-1} \, \phi)^{-1} \, |\nabla \omega| ^3 \, , \qquad           \label{cubic Lagrangian}
\eeq
where we have used the abbreviation
\beq |\nabla \omega| :=   \left|\partial_{\nu}\omega\, \partial^{\nu} \omega \right| ^{\frac{1}{2}}
\eeq 
($| \dots|$ absolute value).\footnote{For timelike $\nabla \omega$ see \citep{Scholz:2015MONDlike}. Here we exclusively deal with the spacelike (or null) case.} 
 $\omega$ is the scale invariant representative of the scalar field introduced in (\ref{tilde omega}), and $\nabla$ its gradient.\footnote{$w(\phi)=-1$ and $ w(\left\| \nabla {\omega}  \right\| =-1$ imply the scale weight $w( L_{\phi 3})=-4$, as it must be for scale invariance of $\mathcal{L}_{\phi 3}$.} 

With $\phi_o$ the (constant) value of $\phi$ in Einstein gauge we  
 introduce the constants 
\beq
a_o := \xi^{-1}\phi_o \,, \qquad  \tilde{a}_o=\frac{a_o}{4} \, . \label{tilde ao}
\eeq
Then  the cubic term of the kinetic Lagrangian in Einstein gauge reads 
\beq L_{\phi 3} \doteq  \frac{2}{3}(8\pi G \, \tilde{a}_o)^{-1} |\nabla \omega|^3 \, .
\eeq 
The {\em dotted equality} sign  ``$\doteq$'', indicates that the respective equation is not scale invariant  but presupposes a special gauge made  clear by the context (similar for  $\dot{\approx}$). Here, as in most cases in this paper, it indicates the {\em Einstein gauge}.
$\tilde{a}_o$ has the dimension of inverse length/time and will play a  role  analogous to the MOND acceleration $a_o \approx [c]\,H$ ($c$ the velocity of light). Coefficients of type $[c]$ and $[\hbar]$ will often be suppressed in the following general considerations. They will be plugged in only in the final step.
The proportional factor in   $\tilde{a}_o = \frac{a_o}{4}$ is chosen such that the WST model with cubic kinematic Lagrangian (WST-3L) acquires a 
MOND-like  phenomenology   in a  weak gravitational field in which the scalar field and the scale connection cannot be neglected.

$\xi^2$ is a typical ``large number'' in the sense  of bridging the gap between the smallest and the largest physically meaningful energy scales  
\[ \xi^2 = \frac{E_{pl}}{a_o [\hbar]}  = \frac{a_o^{-1} [c]}{L_{pl}}   \sim 10^{63} \,. 
\]

To the  kinetic Lagrangian of the scalar field
a potential term is added. It must be of order 4 to provide for scale invariance of the density (eq. (\ref{modified Hilbert term}):
\beq L_{V4}= -\frac{\tilde{\lambda}}{4} \phi^4 
\eeq 
The  absolute value of the corresponding  energy density   can be read off from
\[ \frac{\tilde{\lambda}}{4} \phi^4   =  (\xi \phi)^2\, \frac{\tilde{\lambda}}{4} (\xi^{-1} \phi)^2 \, ;
\]
in Einstein gauge it is constant. With $\lambda = \frac{\tilde{\lambda}}{36} $  it becomes
\beq (\xi \phi)^{-2}|L_{V4}|=  \frac{\tilde{\lambda}}{4} (\xi^{-1} \phi)^2 \doteq \frac{\lambda}{4}\, 36\, a_o^2 \doteq   \frac{\lambda}{4} H^2 \label{H^2}
\eeq 
 comparable to the cosmological constant term $\Lambda = 3\, \Omega_{\Lambda}\, H^2$ of the standard approach. 

Variation of the Lagrangian leads to the dynamical equations of WST, the  Einstein equation and the scalar field equation. The {\em scale invariant} Einstein equation is\footnote{This means that not only the equation but all its constitutive (additive) terms are scale invariant. In particular the Ricci tensor $Ric$ of Weyl geometry is scale invariant because the affine connection of the Weyl metric is.}
\beq Ric-\frac{R}{2}g = (\xi \phi)^{-2} T^{(m)} + \Theta \, , \label{scale invariant Einstein equation}
\eeq 
where $g$ denotes the whole collection of metrical coefficients and $T^{(m)}$ the energy tensor of matter (\ref{matter energy tensor}). 
The scalar field contributes to the total  energy momentum with two terms, $\Theta=\Theta^{(I)}+\Theta^{(II)}$, the first of which is proportional to the metric (thus formally similar to a vacuum energy tensor):\footnote{ See, e.g. \citep{Drechsler/Tann}, \citep[pp. 96ff.]{Blagojevic:Gravitation}.}
\beqarr
\Theta^{(I)} &=& \phi^{-2} \left(  - D_{\lambda}D^{\lambda}\phi^2  +   \xi^{-2}(L_{V4} +  L_{\phi}) \right)g   \,  \label{ThetaI}\\
\Theta^{(II)}_{\mu \nu} &=& \phi^{-2} \left( D_{\mu}D_{\nu}\phi^2 
- 2  \xi^{-2} \frac{\partial L_{\phi}}{\partial g^{\mu \nu}} \right) \,   \qquad    \label{ThetaII}
\eeqarr

Varying with regard to $\phi$ gives the scalar field equation. Subtracting the trace of the Einstein equation  for a conformally coupled $L_{\phi 2}$ term ($\alpha = - 6 \xi^2$)   strongly simplifies it and introduces  the trace of the matter tensor into the scalar field equation.
In Einstein gauge, with $g$ the Riemannian component of the metric,  it can  be written in terms of the covariant derivative $_g\hspace{-0.15em}\nabla$ with regard to $g$ (Levi-Civita connection in Einstein gauge) as\footnote{\citep[pp. 15f., sec. 7.2, postprint version  arXive v4]{Scholz:2015MONDlike}} 
\beq   \, _g\hspace{-0.2em}\nabla_{\nu}(|\nabla \omega | \partial^{\nu} \omega) \doteq - 4 \pi G \, \tilde{a}_o\, tr\, T^{(m)} \, .
\eeq
If we introduce the corresponding Riemannian covariant operator 
\beq _g\hspace{-0.12em}\square_M\, \omega = \, _g\hspace{-0.2em}\nabla_{\nu}(|\nabla \omega | \partial^{\nu} \omega) =\left(\partial_{\nu}|\nabla \omega | \, \partial^{\nu} \omega + | \nabla \omega | \, _g\hspace{-0.2em}\nabla_{\nu} \, \partial^{\nu} \omega   \right) \, , \label{covariant Milgrom op}
\eeq 
the scalar field equation for a fluid with matter density $\rho_m$ and pressure $p_m$ simplifies to the covariant {\em Milgrom equation}
\beq   _g\hspace{-0.12em}\square_M\, \omega  \doteq 4 \pi G \, \tilde{a}_o\, (\rho_m - 3p_m) \, .   \label{Milgrom equation}
\eeq 
In this derivation, with $tr\, T^{(m)}$ entering by subtracting the trace of the Einstein equation, a sign choice like in (\ref{Bekenstein's (6)})  leads to the wrong sign on the r.h.s of the Milgrom equation.   By obvious reasons (\ref{covariant Milgrom op}) will be called the {\em covariant Milgrom operator}.  In the static weak field static case $\omega$ does not depend on the time coordinate. Moreover with $g_{\mu \nu} \approx \eta_{\mu \nu}$, the expression $\nabla_{\nu} (| \nabla \omega | \partial^{\nu}\omega)$ turns into the nonlinear Laplace operator  $\nabla \cdot (| \nabla \omega | \nabla \omega)$ of the MOND theory   with Euclidean scalar product $\cdot$ and norm $| \ldots |$.  

In the general case we have to complement (\ref{Milgrom equation}) with the Einstein equation in Einstein gauge  
\beq Ric-\frac{R}{2}g \doteq 8 \pi G\,  T^{(m)} + \Theta \, , \label{Einstein equation}
\eeq 

In vacuum, the trivial scalar field  $\omega=const$  is a basic solution of (\ref{Milgrom equation}). Then WST  reduces to Einstein gravity.
In particular, the Schwarzschild and the Schwarzschild-de Sitter solutions of Einstein gravity are special (degenerate) solutions of WST-3L equations for  $\frac{\lambda}{4}= 0$ or $\frac{\lambda}{4}\approx 6$, respectively. In fact, they solve (\ref{Einstein equation}), (\ref{Milgrom equation}) for  $\phi \doteq const$ in Riemann gauge, i.e. in the case of Einstein gauge equal to  Riemann gauge $(\hat{g}, \hat{\varphi}) = (\tilde{g}, 0)$. The Riemannian component of the  metric ($\tilde{g}=\hat{g}=:g$) is given by
\beq ds^2 = - (1-\frac{2M}{r} -\kappa\, r^2)dt^2 + (1-\frac{2M}{r} -\kappa\, r^2)^{-1}dr^2 + r^2(dx_2^2 + \sin^2 x_2\, dx_3)^2 \, . \label{Schwarzschild-de Sitter}
\eeq
Then
$ Ric - \frac{R}{2}g = - 3\kappa\, g $ and
$ \Theta = \Theta^{(I)}= -\frac{\lambda}{4}\beta^2 \tilde{a_o}^2 \, g  $.
Therefore the Einstein equation is satisfied for $3 \kappa = \frac{\lambda}{4} \beta^2 \, \tilde{a_o}^2$, i.e.  $\kappa \approx 2H^2$ for $\frac{\lambda}{4} \approx 6$ and $ \beta  \approx 100$. 
We see that in the case of   a { negligible}  { Weylian scale connection} the classical (non-homogeneous) { point-symmetric solutions of Einstein gravity} are valid also for the dynamics of WST. This implies  that {\em in the case of a negligible scale connection}  Newton dynamics  is an effective approximation for point symmetric solutions of WST  (in Einstein gauge).

 In order to make such a type of Einstein limit compatible with our Lagrangian,  a suppression of the $L_{\phi 3}$-term for sufficiently large accelerations $a_R$  of (\ref{a_R}) is necessary.  Following the example of  ordinary MOND theories, one might   be tempted to plug  a factor $\tilde{f}(\frac{a_o}{|a_R|})$  with a function $\tilde{f}$ such that $\tilde{f}(y) \sim 1$ for $y>0.01$  and $\tilde{f}(y) \sim 0$ for $y\ll1$ into   the r.h.s expression of  (\ref{cubic Lagrangian}) for  $L_{\phi_3}$. But such a choice would have the blemish of a coordinate dependent argument of the function.
A better alternative is provided
by the {\em hypothesis} that the scalar field inhomogeneities are
suppressed if any of the sectional curvatures $\kappa$ (with respect to
the Riemannian component of the metric in Einstein gauge)
surpasses a certain threshold, e.g., $\kappa \geq (10^9 a_o[c^{-2}])^2$. In
the next section we investigate the case of a non-negligible
scale connection. A more detailed discussion of the transition
between the two domains has to be left open for  another
occasion.


\subsection{\small A WST approach with MOND-like phenomenology \label{subsection MOND}}
If the conditions for the weak field approximation $(\ref{weak field})$ are given, it is possible to identify a   {\em MOND regime} as a region in which the  Newton acceleration $a_N$ is smaller than $a_o$ (here $a_N$ can be identified with $a_R$ in   (\ref{Newton limit})). 
Then the scalar field equation (\ref{Milgrom equation}) reduces, in reliable approximation, to 
\beq  \nabla \cdot (| \nabla \omega | \nabla \omega)   \doteq  -    4 \pi G \, \tilde{a}_o \;  tr\, T^{(m)} \, , \label{MOND approx SF equation}
\eeq 
with the Euclidean $\nabla$-operator.  
We call this a {\em MOND approximation}.
For pressure-less matter with energy density $\rho_m$  we get
\beq \nabla \cdot (| \nabla \omega | \nabla \omega)   \doteq   4 \pi G\, \tilde{a}_o\, \rho_m \, . \label{non-linear Poisson equ}
\eeq 
That is  similar to the AQUAL approach.\footnote{ \citep{Bekenstein/Milgrom:1984}, \citep{Bekenstein:2004}.}
Note that only  the {\em matter} energy momentum tensor,  {\em  without} the {\em scalar field energy density}, appears on the r.h.s. of (\ref{MOND approx SF equation}).

Straight forward verification shows that, independent of  symmetry conditions,  
a solution of (\ref{non-linear Poisson equ}) is given by $\omega$ with a gradient  $\nabla \omega = - a_{\varphi}$ such that 
\beq   a_{\varphi} = \sqrt{\frac{\tilde{a}_o}{|a_N|}}\, a_N
=     \sqrt{\tilde{a}_o|a_N|}\, \frac{a_N}{| a_N|}   \,  ,\label{solution non-linear Poisson equ}
\eeq 
where $a_N$ denotes  the  Newton acceleration  of the given mass density, 
\beq \nabla^2 \Phi_N=4\pi G\, \rho_m \, \qquad a_N=- \nabla \Phi_N  \label{a-N Phi-N}
\eeq
(calculations in the approximating Euclidean space with norm $| \ldots |$).
The solution of the non-linear Poisson equation (\ref{non-linear Poisson equ}) is much simpler  than one might expect at a first glance: In a first step the linear Poisson equation of the Newton theory is to be solved, then an algebraic transformation of type (\ref{solution non-linear Poisson equ}) leads to the acceleration due to the solution of the non-linear partial differential equation (\ref{non-linear Poisson equ}).  In fact, $a_{\varphi}$ has the form of the deep MOND acceleration of the ordinary MOND theory (but with a different constant $\tilde{a}_o$).\footnote{In the terminology of the MOND community, the MOND approximation of  WST-3L behaves like a special case of a QMOND theory \citep[pp. 46ff.]{Famaey/McGaugh:MOND}.}

This is only the most immediate modification of  Newton  gravity. In (\ref{Einstein equation})  there is also the additional term  of the {\em energy density due to the scalar field}, $\rho_{s\hspace{-0.1em}f } = (8 \pi\, G)^{-1}\Theta_{oo}$. It modifies the r.h.s. of the Newton limit of Einstein gravity.\footnote{In contrast  $\rho_{s\hspace{-0.1em}f }$ does not enter the r.h.s. of the scalar field equation (\ref{Milgrom equation}), and therefore does {\em not enter} the r.h.s of (\ref{a-N Phi-N}).}
Neglecting contributions at the order of  magnitude of cosmological terms ($\sim H^2$) and of $|\nabla \omega|^2$), the energy density of the scalar field in Einstein gauge simplifies to
\beq \rho_{s\hspace{-0.1em}f } \;  \dot{\approx} \;  (4\pi G)^{-1} \nabla^2 \omega\, , \label{SF energy density}
\eeq 
where Latin indices $j, k$ \ldots refer to space coordinates only (see appendix 5.1).
 
That is equal to  the value of the phantom energy density corresponding to the acceleration of the scale connection (\ref{phantom energy of scale connection}). The total ``anomalous'' additive acceleration  (in comparison to Newton gravity) is therefore 
 \beq a_{add} = a_{\varphi} + a_{sf}= 2 a_{\varphi}\, . \label{a-add}
\eeq
In the central symmetric case
\beq |a_{add}| = 2 \frac{\sqrt{GM(r)\, \tilde{a}_o}}{r} \,.
\eeq 
In the case of  $a_N \ll a_{add}$ this leads to MOND-like phenomenology in the deep MOND regime if
 \beq
\tilde{a}_o= \frac{a_o}{4} \approx \frac{1}{24}\, H\, [c] \approx 2.4 \cdot 10^{-9}\, cm\, s^{-2} \, . \label{a-tilde}
\eeq
Because of (\ref{solution non-linear Poisson equ}) the total acceleration $a$ is then
\beq a= a_N+a_{add} =  a_N \left(1+ \sqrt{\frac{a_o}{| a_N |}} \right) \,, \qquad |a_{add}| = \sqrt{a_o | a_N |} \, . \label{total acceleration}
\eeq

This raises the question of the Newtonian limit. (\ref{solution non-linear Poisson equ})
implies  $|a_{\varphi}| \ll |a_N |$ in regions where   $|a_N| \gg a_o \,(> \tilde{a}_o)$.  Therefore $a_{\varphi}$  can  effectively  be neglected in the case of `large' values of $|a_N|$ derived from (\ref{a-N Phi-N}). Assuming the hypothesis at the end of section \ref{subsection WST-3L} (or some equivalent), the Newton approximation is   reliable in WST gravity,   irrespective of the  question of how to characterize the  transition between the MOND and the Newton approximation. Here we shall consider the MOND approximation  {\em in  an ``upper transition'' regime} only, where  roughly $|a_N| \leq 10^2 a_o$.\footnote{One might speak of the  {\em upper transition regime} for $a_o\leq |a_N|\leq 100\, a_o$, of the {\em MOND regime} if $|a_N|\leq a_o$ and of  the {\em deep MOND regime} for, let us say, $|a_N|\leq 10^{-2}a_o$     \citep[sec. 7.3]{Scholz:2015MONDlike}.  We cannot claim knowledge on the  ``lower'' transition regime  with $100\, a_o < |a_N| $ but not yet $a_o\ll |a_N|$ (however $\ll$ may be specified); see end of section \ref{subsection WST-3L}. \label{fn transition regime}}

For centrally symmetric mass distributions $\rho(r)$ with mass $M(r)$ integrated up to $r$  (where $r = |y|$ denotes the Euclidean distance from the symmetry center,  $y=(y_1,y_2,y_3)$ the coordinates of the approximating Euclidean space) this implies
\beq
a_{\varphi} = - \nabla \omega  \,  \dot{\approx}   - \sqrt{GM(r)\, \tilde{a}_o}\frac{y}{|y|^2} \, , \qquad | a_{\varphi}| =  \frac{\sqrt{GM(r)\, \tilde{a}_o}}{r} \, . \label{a-phi point symmetric} \eeq
Then the phantom energy density (\ref{phantom energy of scale connection}) becomes\footnote{For $ds^2 = dr^2 + r^2 (d\theta^2 + \sin^2 \theta\, d\vartheta^2)$ the crucial affine connection components are $\Gamma_{11}^1 =0,\, \Gamma_{2 1}^2 = \Gamma_{3 1}^3 = r^{-1}$.}
\[\rho_{ph} = (4 \pi G)^{-1} \frac{\sqrt{GM(r)\, \tilde{a}_o}}{r^2} \, ;
\] 
and also
\beq \rho_{s\hspace{-0.1em}f } \;  \dot{\approx} \;  (4 \pi G)^{-1} \frac{\sqrt{GM(r)\, \tilde{a}_o}}{r^2}
\, . \eeq

\subsection{\small Comparison with usual MOND theories  \label{subsection Comparison}}
We can now  compare our approach with  other  models of the MOND family. Simply adding a deep MOND term to the Newton acceleration of a point  mass  is unusual.  M. Milgrom rather considered a multiplicative relation between the MOND acceleration $a$ and the Newton acceleration $a_N$ by a kind of `dielectric analogy',
\beq
a_N = \mu(\frac{a}{a_o}) \, a \; , \qquad \mbox{with} \quad \mu(x)\longrightarrow  \left\{{ 1 \quad  \;  \mbox{for}\; x \to \infty} \atop { x \quad \; \; \mbox{for} \; x\to 0 \; ,} \right. \label{mu function}
\eeq 
or the other way round\footnote{Here $\mu(x)\to x$ means $\mu(x)-x = \mathcal{O}(x)$, i.e. $\frac{\mu(x)-x}{x}$ remains bounded for $x \to 0$. Cf. \citep[51f.]{Famaey/McGaugh:MOND}}
\beq
a = \nu (\frac{a_N}{a_o}) \, a_N \; ,  \qquad \mbox{with} \quad \nu(y) \longrightarrow 
\left\{{ 1 \hspace{2em} \mbox{for} \; y \to \infty} 
\atop {  y^{-\frac{1}{2}} \; \;   \mbox{for} \; y\to 0 \; . } \right. \label{nu function}
\eeq  
From this point of view our acceleration (\ref{total acceleration}) is specified by a well defined transition functions 
\beq \mu_w(x) = 1+ \frac{1-\sqrt{1+4x}}{2x} \qquad \mbox{and}
\quad 
\nu_w(y) = 1+ y^{-\frac{1}{2}} \; . \label{mu-w}
\eeq
One has to keep in mind, however, that our {\em transition functions} $\mu, \nu$ {\em are  reliable only  in the MOND regime and the upper transitional regime} (roughly $a_N \leq 10^{2} a_o$). They cannot be used for discussing the Newtonian limit.\footnote{See fn. \ref{fn transition regime} and the text above it.}
It will be important to see how they behave in the light of empirical data, in particular galactic rotation curves and cluster dynamics.

In the MOND literature the amount of a (hypothetical) mass which  in Newton dynamics would produce the same effects as the respective MOND correction $a_{add}$ is called {\em phantom mass} $M_{ph}$.
  For any member of the MOND family the additional acceleration can be expressed  by the modified  transition function $ \tilde{\nu}= \nu -1 $ with $\nu$ like in (\ref{nu function})
\beq a_{add}= \tilde{\nu}\left(\frac{|a_N|}{a_o}\right)\, a_N \, .
\eeq 
The {\em phantom mass density} $\rho_{ph}$ 
 attributed  to the the potential $\Phi_{ph}$ satisfies $4 \pi G\, \rho_{ph} = \nabla^2 \Phi_{ph} $ and $\nabla \Phi_{ph}= - a_{add}$. A short calculation shows that it may be expressed as
\beq  \rho_{ph}=  \tilde{\nu}\left(\frac{|a_N|}{a_o}\right)\, \rho_m 
- (4\pi G\, a_o)^{-1} \tilde{\nu}'\left(\frac{|a_N|}{a_o}\right)\, (\nabla| a_N |) \cdot a_N  \, .  \label{rho-ph}
\eeq
It consists of a contribution proportional to $\rho_m$ with factor $\tilde{\nu}$, which dominates in regions of ordinary matter, and a term derived from the gradient of $|a_N|$ dominating in the ``vacuum'' (where however scalar field energy is present).
For the Weyl geometric model with $\tilde{\nu}_w(y)=y^{-\frac{1}{2}}, \, \tilde{\nu}_w'(y)=-\frac{1}{2}y^{-\frac{3}{2}}$ $\rho_{ph}$ turns into:
\beqarr  \rho_{t}&=& \left(\frac{a_o}{|a_N|}\right)^{\frac{1}{2}}\, \rho_m + (8\pi G)^{-1}\left(\frac{a_o}{|a_N|}\right)^{\frac{1}{2}}\, \nabla(| a_N |)\cdot \frac{a_N}{|a_N|} \\
\rho_{s\hspace{-0.1em}f } &=& \rho_{ph}  =\frac{1}{2} \rho_{t} \, \label{rho-ph Weyl}
\eeqarr
The first expression of (\ref{rho-ph Weyl}) is  compatible with (\ref{SF energy density}).

 In our case it would be utterly wrong to consider  the whole of  $\rho_{t}$ as ``phantom energy''. Half of it are due to the scalar field energy density, the {\em scalar field halo} $\rho_{s\hspace{-0.1em}f }$, and expresses a {\em true energy density}. This energy density appears  on the right hand side of the Einstein equation (\ref{Einstein equation}) and the Newtonian Poisson equation as  its weak field, static limit. It is decisive for {\em lensing} effects of the additional acceleration. 
The other half, $\rho_{ph}$, is phantom, i.e. a fictitious mass density producing the same acceleration  as the Weylian scale connection  (\ref{phantom energy of scale connection}). Only for the sake of comparison with other MOND models we may speak of  $\rho_{t}$ as some kind of {\em gross} phantom energy,in contrast to the ``net'' phantom energy   $\rho_{ph\,1}$.

We have to distinguish between the influence of the additional structure, scalar field and scale connection, on light rays and on (low velocity) trajectories of mass particles. Bending of light rays is influenced by the scalar field halo only, the acceleration of  massive particles with velocities far below $c$ by   the scalar field halo {\em and} the scale connection.\footnote{That may look like bad news for explaining lensing at clusters and microlensing at substructures. But  the particular transition function seems to compensate much of this effect.} 

Also in another respect our theory differs from the usual MOND approaches. In MOND {\em external acceleration} fields of a system under consideration are  difficult to handle. 
In WST, like in GR,  a  freely falling (small) system does not feel the external acceleration field if it is   sufficiently small, relative to the inhomogeneities of the external gravitational field, for neglecting tidal forces. In this sense, the  external acceleration problem does not arise in the  WST MOND approximation (\ref{MOND approx SF equation}).

Another important consequence follows: the scalar field energy formed around a freely falling subsystem of a larger gravitating system,  calculated in the MOND approximation of the freely falling subsystem, contributes to  the r.h.s. of the Einstein equation of any other subsystem (in relative motion) and also to that of a superordinate larger system.\footnote{In principle that presupposes that the whole energy momentum tensor (\ref{ThetaI}, \ref{ThetaII}) (and its system dependent representation) is considered. For slow motions and weak field approximation a  superposition of energy densities like in Newton dynamics seems legitimate.} This has to be taken into account for modelling the dynamics of clusters of galaxies.


\subsection{\small Short  resum\'e  \label{subsection resumee}}
We have  derived the most salient features of the Weyl geometric MOND approximation (WST MOND) and are prepared for a comparison with empirical data. Before we do so, it may be worthwhile to collect the results  which are necessary for applying it to real constellations   in a short survey. 

Consider a gravitating system which  in the Newton approximation of Einstein gravity is described by the baryonic matter density $\rho_{m}$, the acceleration $a_{m}$ and potential $\Phi_{m}$ with
\beq \nabla^2\Phi_{m}= 4\pi G \rho_{m} \, , \qquad a_{m}=-\nabla \Phi_{m}.
\eeq
The modification due to  WST MOND leads to an additional acceleration $a_{add}$ with the following features: 
\begin{itemize}
	\item[\qquad (i)] The total acceleration $a$ is $ a= a_{m}+ a_{add}$ with
\beq a = a_{m}\left(1+\left(\frac{a_o}{|a_{m}|}\right)^{-\frac{1}{2}}\right) = \nu\left(\frac{|a_{m}|}{a_o}\right)a_{m}\, , \label{equ (i)}
\eeq 
where $\nu(y)=1+y^{-\frac{1}{2}}$. For $a_m \gg a_o$ the Newton approximation applies. (\ref{equ (i)}) holds  for $|a_m| \leq 10^2 a_o$ only (``upper'' transition regime). No information can be drawn from it for  $|a_m|$ larger but not yet $\gg a_o$  (the ``lower'' transition regime). 
\item[(ii)] The ``reciprocal'' transformation function  defined by  $a_{m}= \mu\left( \frac{|a|}{|a_{m}|} \right)a $  is 
\beq \mu(x) = 1+ \frac{1-\sqrt{1+4x}}{2x}\,.
\eeq
		\item[(iii)]  $a_{add}$ consists of two components $a_{add}=a_{\varphi}+a_{s\hspace{-0.1em}f }= 2 a_{\varphi}$. The first one is derived from a potential $\omega$ satisfying the  non-linear Poisson equation
		\beq \nabla \cdot (|\nabla \omega | \nabla \omega)= \pi G a_o \, \rho_{m} \, , \qquad a_{\varphi} = - \nabla \omega\, .
				\eeq
	\item[(iv)] The second one, $a_{s\hspace{-0.1em}f }$, can be understood as a Newton acceleration due to the energy density $\rho_{s\hspace{-0.1em}f }$ of a scalar field (part of the modified gravitational structure). Its potential satisfies a Newtonian Poisson equation. It satisfies
		\beq - \nabla a_{s\hspace{-0.1em}f }= 4 \pi G \rho_{s\hspace{-0.1em}f }
	\eeq
	with energy density 
	\beq  \rho_{s\hspace{-0.1em}f } = \frac{1}{2} \left(\frac{a_o}{|a_m|}\right)^{\frac{1}{2}}\, \left( \rho_m + (8\pi G)^{-1} \nabla(| a_m |)\cdot \frac{a_m}{|a_m|}\right)  \label{scalar field halo}
	\eeq 
$\rho_{s\hspace{-0.1em}f }$ is part of the energy-momentum tensor of the scalar field $\phi$ and in this sense ``real'' rather than phantom. 
		\item[(v)]  $a_{\varphi}$  is formally  derivable in  Newton dynamics   from a fictitious energy density 
			\beq  \rho_{ph} =   \rho_{s\hspace{-0.1em}f } \, . \label{phantom halo}
			\eeq 
			$\rho_{ph}$ is the net phantom energy of WST MOND. For comparison with other models of the MOND family one may like to consider $\rho_{ph} + \rho_{s\hspace{-0.1em}f }= \rho_{t}$ as a kind of ``gross phantom energy'' (although the larger part of it is real). It is transparent rather than ``dark'' (see  (\ref{transparent matter halo}) below).
			\item[(vi)] (i) -- (v) are reliable approximations also for  small (local) gravitating systems freely falling in a larger gravitating system, if    $a_{m}\lessapprox 10^{2}\, a_o$ in the local system. The subsystem can be considered as ``small'' with respect to the super-system, if tidal forces of the super-system can be neglected. In hierarchical systems like galaxy clusters the energy density contributions $\rho_{s\hspace{-0.1em}f }$ of the subsystems and the super-system (calculated in different reference coordinate systems) add up to the total energy density of the scalar field, if the velocities of the subsystems relative to the barycenter of the super-system are small. This is a crucial difference between WST MOND and  ordinary MOND theories.
			If one likes, $\rho_{s\hspace{-0.1em}f }$ can be considered as the ``dark matter'' component of WST MOND; although as the energy density of the scalar field it is not constituted by the usual (hypothetical) quantum particles (WIMPs, axions etc.). To demarcate this difference it might better be called {\em transparent} matter/energy of
			WST.
			\item[(vii)] Gravitational lensing is due to the scalar field energy density only ($\rho_{s\hspace{-0.1em}f } $), while the dynamics of WST corresponds to the total phantom density  ($\rho_t=\rho_{s\hspace{-0.1em}f } + \rho_{ph}$). It remains to be seen whether such a difference is in agreement with observations.
\end{itemize}


\section{ Halo model for clusters of galaxies \label{section halos}}

\subsection{\small Cluster models for baryonic mass (hot gas and stars) \label{subsection Cluster model}}
In the astronomical literature, the density profile of hot  gas and  (smeared) star/galaxy matter in a galaxy cluster is often described by a  centrally symmetric profile of the following form: 
\beq \rho(r)= \rho_o \left( 1+ \left( \frac{r}{r_c} \right)^2 \right)^{-\frac{3}{2} \beta} \,   \label{beta model}
\eeq 
 $\beta$ is the ratio of the specific energies of the galaxies and the gas, $\rho_o$  the central density and  $r_c$ is the core radius \citep{Sanders:1999,Sanders:2003,Reiprich:Diss}.\footnote{$r_c$ is the distance from the cluster center at which the projected galaxy density is half  the central density $\rho_o$.}  
(\ref{beta model}) is called a  {\em $\beta$-model} for the mass distribution. For our test we assume   density models for the gas mass $\rho_{gas}(r)$ and  for the galaxy mass $\rho_{star}(r)$ with the same form  parameters $\beta$ and $r_c$.
We thus work with an idealized model using proportional density profiles for the hot gas and for the galaxies with parameters $\beta, r_c$ determined from observations of  the hot gas.   The central densities $\rho_o$ can, in principle,  be determined from mass data for gas $M_{gas}(r_1)$, respectively stars $M_{star}(r_1)$,  at a given distance $r_1$. The empirical determination of $M_{gas}(r_1)$ and $M_{star}(r_1)$ from directly observable quantities is a subtle question; it  will be discussed  in section 
\ref{subsection empirical data I}. 

Large scale gravitational effects on the cluster level are often modelled in the Newton approximation of Einstein gravity with baryonic matter and an assumed dark matter halo which is inferred from its gravitational effects (in Einstein/Newton gravity).  Another, minority, approach in the literature  works with an evaluation of the data in a MOND limit of the most well known relativistic MOND theory TeVeS. R. Sanders is one of its protagonists;  he  concludes that in this approach a much smaller amount of unseen matter has to be assumed in addition to the baryonic mass. Its value is  consistent with the hypothesis of a  halo of sterile neutrinos, concentrated about the cluster center \citep{Sanders:2003}. Here we want to explore the feasibility of the WST approach, in particular regarding the question of how much unseen matter has to be added to the gravitational effects of the model in order to reproduce (``predict'') the observed accelerations, respectively their  measurable effects. 

\subsection{\small Two contributions to the scalar field energy in clusters of galaxies}
\label{subsection two contributions}}
The mass distribution of the hot gas $ \rho_{gas}$  in a  galaxy cluster is described by a $\beta$-profile (\ref{beta model}) in a  locally static coordinate system with origin at the barycenter of the cluster. The averaged star mass will  be represented by a continuous distribution $\rho_{star}$ of a $\beta$-profile  with the same parameters $\beta, r_c$, but with a different value of $\rho_o$. Both together form a continuity model of the   baryonic  mass distribution $\rho_{bar}=\rho_{gas}+\rho_{star}$.
Estimates  show that Newtonian gravitational accelerations induced by $\rho_{bar}$ (far away from mass concentrations stars, galaxies, and galactic centers) are below $10^2\cdot a_o$. They are  small enough for  allowing to in a weak field static approximation with the scalar field equation in the MOND approximation (\ref{non-linear Poisson equ}). The resulting contribution to the scalar field energy will be called $\rho_{s\hspace{-0.1em}f \, 1}$. The additional acceleration of the Weyl geometrical scale connection (\ref{phantom halo}) can be expressed in terms of a phantom mass density which will be called $\rho_{ph \, 1}$.

In this    first approximation  the star mass is approximated on a par with the hot gas, i.e., it is described by its continuously smeared out mean density. But stars are agglomerated in galaxies  which form freely falling subsystems of the cluster with considerable inter-spaces in the super-system (the cluster).  For each subsystem a locally static coordinate system with origin at the respective galactic center can be chosen. In this system the local inhomogeneities of star mass distribution  in the cluster, and the resulting inhomogeneities of the gravitational field, in the vicinity of of the galaxy can be calculated.  
On the galaxy level the MOND theory has proven effective for modelling  gravitational effects deviating from Einstein and Newton gravity without assuming real dark matter \citep{Famaey/McGaugh:MOND}. Although we expect that the MOND approximation of WST gravity shows similar features,  this is not the point in the present investigation. 

Here we are interested in the neighbouring regions of galaxies as subsystems of their respective cluster. 
 {\em These  subsystems form scalar field halos of their own} which contain real energy  (different from the classical MOND theory which leads to phantom halos  only).  In the framework of the present approach, the scalar field halos in the neighbourhood of each galaxy contribute to the  energy density which adds up  globally, i.e. on the cluster level, to a component of scalar field energy  $\rho_{s\hspace{-0.1em}f \, 2}$ which has been suppressed in the first continuity approximation of the total baryonic mass. 
In a second step we therefore   determine this component approximately and add it to the total the scalar field halo.

\subsection{\small Scalar field and phantom halos  $\rho_{s\hspace{-0.1em}f \, 1}$,   $ \rho_{ph \, 1}$  in the cluster-barycentric MOND approximation\label{subsection baryonic halos}}

The baryonic mass up to radius $r$,
\beq M_{bar}(r)=4\pi \int_0^r \rho_{bar}(u)\, u^2 du \, , \label{Mbar}
\eeq 
determines the Newton acceleration $a_{bar}= G \frac{M(r)}{r^2}$ due to the  total baryonic mass. The densities of the scalar field halo $ \rho_{s\hspace{-0.1em}f \, 1} $ and the phantom halo of WST MOND follow from (\ref{scalar field halo}), (\ref{phantom halo}). They are
\beqarr  \rho_{s\hspace{-0.1em}f \, 1} &=&\frac{1}{2} \left(\frac{a_o}{|a_{bar}|}\right)^{\frac{1}{2}}\, \left( \rho_m + (8\pi G)^{-1} \nabla(| a_{bar} |)\cdot \frac{a_{bar}}{|a_{bar}|}\right)\, ,  \label{scalar field halo baryonic matter} \\
 \rho_{ph \, 1} &=&  \rho_{s\hspace{-0.1em}f \,1} \, .\label{phantom halo baryonic matter}
\eeqarr 
The respective masses of the halos $M_{s\hspace{-0.1em}f \, 1}, \, M_{ph \, 1}$ arise from integration.

\subsection{\small  Scalar field halos of  galaxies in their respective galacto-centric MOND approximations  \label{subsection scalar field halo galaxies}}
As already indicated in section \ref{subsection two contributions} (\ref{scalar field halo baryonic matter}) and (\ref{phantom halo baryonic matter})   do  not  make allowance for the fact that the  star matter forms a discrete  structure of an ensemble of galaxies {\em each of which is falling freely} in the inertio-gravitational field of the super-system (hot gas and other galaxies). Every galaxy possesses a local MOND approximation with regard to its own barycentric static reference system.  The acceleration $a_{bar}$ of the super-system (with respect to the barycenter rest system of the hot gas) is  transformed away in each of the local MOND approximations. The latter leads to a galactic {\em  scalar field halo}  which {\em  persists under changes of reference systems} with small, i.e. non-relativistic, relative velocities. It contributes to the total energy of the scalar field, calculated in the cluster barycentric system. (Of course this is {\em not} the case for the phantom halo of the single galaxies.) 
 In principle, we have to add up all these effects to  a scalar field energy density  $\rho_{s\hspace{-0.1em}f \,2} $ in order to fill in this lacunae. But an exact calculation would have to solve a highly non-trivial  $N$-body problem for the motion of the galaxies.

The experience with the calculation of the combined MOND halos of stars inside galaxies shows that a resolution of the star matter inside galaxies down to individual galaxies is not necessary to achieve good results. In the outer region of galaxies a continuity model for the distribution of star matter in the galactic disk gives reliable approximations for the MOND acceleration.
Similarly we want to check  whether  also here a  continuity model for the system of  galaxies alone, abstracting from the gas mass,  leads to an acceptable approximation for $\rho_{s\hspace{-0.1em}f \,2}$. For this calculation, the gas mass has to be omitted because the galaxies are  falling freely in the outer field of the cluster; the gravitational potential of the latter does not enter the local MOND approximation of the galaxies.

Using   (\ref{scalar field halo})  again we get for the second (inhomogeneity) component of the scalar field energy 
\beq \rho_{s\hspace{-0.1em}f \,2} = \frac{1}{2} \left(\frac{a_o}{|a_{star}|}\right)^{\frac{1}{2}}\, \left( \rho_{star} + (8\pi G)^{-1} \nabla(| a_{star} |)\cdot \frac{a_{star}}{|a_{star}|}\right)\, ,  \label{scalar field halo galaxies} 
\eeq 
with $|a_{star}(r)| = G\frac{M_{star}(r)}{r^2}$ and $M_{star}(r)$ the integral analogous to (\ref{Mbar}) for the star density $\rho_{star}$.
 
We finally arrive at  a  halo model for galaxy clusters constituted by the  components $ \rho_{s\hspace{-0.1em}f \,1}, \rho_{s\hspace{-0.1em}f \,2},  \rho_{ph \, 1}$. All of them are   determined by the two component baryonic profile of the cluster.

\subsection{\small  A three-component halo model for  clusters of galaxies \label{subsection 3 components}}
In addition to  the Newtonian gravitational effects of the {\em  baryonic mass density}
\beq \rho_{bar}= \rho_{gas} + \rho_{star}\, , \label{rho bar}
\eeq  
modeled by a $\beta$-model of type (\ref{beta model}), the WST MOND approach predicts  accelerations generated by  the {\em  scalar field  halos} 
 (\ref{scalar field halo baryonic matter}), (\ref{scalar field halo galaxies}). Because of their  small values  their combined effect can be approximated by a   linear   superposition in the barycentric reference system  of the cluster.\footnote{Because of slow (i.e., non-relativistic) relative velocities of the galaxies   the energy densities of their respective scalar field halos  can be taken over to the cluster barycentric reference system.}   
 \beq
 \rho_{sf} \approx  \rho_{s\hspace{-0.1em}f \,1} +  \rho_{s\hspace{-0.1em}f \,2} \, .\label{total scalar field halo}
\eeq
Moreover, there arises 
an acceleration $a_{\varphi}$  due to the scale connection in the barycentric rest system of the cluster (\ref{a-N Phi-N}). Its gravitational effects are representable by 
 the fictitious  (net)  {\em  phantom halo} $\rho_{ph\,1}$  of (\ref{phantom halo baryonic matter})
\[  \rho_{ph \, 1} =  \rho_{s\hspace{-0.1em}f \,1} \, .
\]  
On the other hand, the  phantom energies of the individual freely falling galaxies do not survive the transformation to the cluster rest system and do not play a role on the cluster level. 

In the usual MOND theories there is no scalar field energy; all additional effects with regard to Newton dynamics may be ascribed to a (fictitious) phantom energy density. Phantom energy densities of single galaxies do not survive the transformation to the cluster barycentric system. In MOND there is therefore no analogy to $ \rho_{s\hspace{-0.1em}f \,2} $; the latter is the {\em crucial distinctive feature} between the two approaches. For a comparison of WST-3L and usual MOND approaches with regard to  galaxy clusters it is  {\em not sufficient to evaluate the difference between the transformation functions} $\mu(x)$ (\ref{mu-w}).

 For an even wider comparison with other approaches  it may be useful to  add  up the scalar field and phantom halos to a kind of  ``dark matter'' halo or, more precisely, to a substitute for the latter.   But one must not forget that in the present model  there is no dark matter in the ordinary sense. Here we only find   a   {\em transparent} halo  made up of  the (real) energy density of the scalar field and the (fictitious) phantom energy density ascribed to the acceleration effects of the scale connection in Einstein gauge (with respect to the cluster barycentric rest system):
\beq  \rho_{t} =  \rho_{s\hspace{-0.1em}f } +  \rho_{ph\,1} \,  \label{transparent matter halo}
\eeq 
From the gravitational lensing point of view, it would be even more appropriate to consider   $ \rho_{s\hspace{-0.1em}f } $ alone as the WST equivalent of a dark matter halo, not forgetting that even the real  halo $  \rho_{s\hspace{-0.1em}f }$ is not due to fermionic particles, but to the scalar field, and thus to the extended  gravitational structure of WST.  From a quantum point of view,  the scalar field  has to be quantized if one wants to search for a (bosonic) particle content of $\rho_{t}$ or $\rho_{s\hspace{-0.1em}f } $. 

The total {\em dynamical mass} of the model (up to some distance $r$ from the center of the cluster) is  
\beq M_{tot}= M_{bar} + M_{t}   , \quad M_{t}=   M_{s\hspace{-0.1em}f }  + M_{ph\,1}, \quad  M_{s\hspace{-0.1em}f }=  M_{s\hspace{-0.1em}f \,1} +   M_{s\hspace{-0.1em}f \,2} \, 
\label{total mass}
\eeq 
with $M_{bar}=   M_{gas }+ M_{star}$. The {\em lensing mass} is a bit smaller,  
\beq M_{lens}= M_{bar} +  M_{s\hspace{-0.1em}f }  \, . 
\label{lensing mass}
\eeq 

Mathematically, the integral of the scalar field energy density to arbitrary distances diverges. Like in the dark matter approach a {\em virial radius} of a cluster may be defined, which roughly delimits the  gravitational binding zone of the cluster. Comparing $\rho_{t}$ (respectively ($\rho_{sf}$)) with the critical energy density  $\rho_{crit}$ of the universe, one may, for example  choose $r_{200}$ with
\beq \rho_{t}(r_{200}) \approx 200\, \rho_{crit}
\eeq 
as a representative of  the virial radius.

Not far beyond the gravitational binding zone of the cluster,  the energy density will have fallen to such a small amount that its centrally symmetric component is inconceivably stronger than the density fluctuations in  the inter-cluster space. To  continue the integration into this region, and beyond, has no physical meaning.   In the long range the energy density of the scalar field approaches the cosmic mean energy value. A physical limit of integration has to be  chosen  close to the virial radius beyond which the gravitational binding structure of the cluster is fading out.

\section{A first comparison with  empirical data  \label{section test}}

\subsection{\small Empirical determination of  mass data for galaxy clusters
\label{subsection empirical data I}}
For  a first empirical exploration we confront  the WST cluster model with recent mass data for 19 clusters obtained  on the background of Einstein/New\-ton gravity and $\Lambda$CDM  by a group of astronomers about Y.-Y. Zhang, T.F. Lagan\'a and T. Reiprich \citep{Reiprich/Zhang_ea:2011}, \citep{Reiprich/Zhang_ea:Corr}.    We use the  form parameters $\beta, r_c$ of the $\beta$-models of these  clusters, published in an earlier study by one author of the group \citep{Reiprich:Diss}. In the present  study we have to take the mass data  and the form parameters essentially at face value. Methodological questions arising from this procedure are discussed in the next subsection. There seem to be sufficient reasons for expecting that the different background theories do not principally invalidate the results thus obtained. Of course, an authoritative empirical study would presuppose an evaluation of observational data on the background of WST itself; it can be done only by astronomers, if they get interested in the present approach.

The studies of Zhang/Lagan\'a/Reiprich e.a. have the great advantage to  build upon three independent observational data sets for determining the  gas mass $M_{gas,500}$, the star mass $M_{\ast,500}$ and the total mass $M_{500}$ (assuming a dark matter explanation for the observed gravitational effects) at the reference distance $r_{500}$. The latter is determined for each cluster at the distance $r_{500}$ from the cluster center at which the total gravitational acceleration indicates a total mass density 500 times the critical density. 
\begin{itemize}
	\item[(i)]   $M_{gas,500}$ has been extracted from X-ray data on the hot intracluster medium (ICM) collected by  XMM Newton and ROSAT. Surface brightness data have been used to infer  an ICM radial electron number density profile, and spectral analysis data gave information on the radial temperature 
	distribution. From that a gas density distribution has been reconstructed and the gas masses at $r_{500}$ by integration.\footnote{An outline of the procedure and literature for more details is given in \citep[3]{Reiprich/Zhang_ea:2011}.}
	\item[(ii)] $M_{\ast,500}$ has been determined from optical  imaging data due to SDSS 7 in two steps. First  the total luminosity of the cluster has been determined by means of  a ``galaxy luminosity function'' (GLF); then the mass is  estimated using mass-to-light ratios depending on the cluster mass. In the last step models of the star development in the respective galaxy, elliptical or spiral, enter. They  depend  on assumptions on an ``initial mass function'' (IMF).  Two possibilities for the IMF (Salpeter versus Kroupa) are considered and compared in  \citep{Reiprich/Zhang_ea:2011}, \citep{Reiprich/Zhang_ea:Corr}. According to the authors the difference of the stellar mass estimate can result in a factor 2  \citep[p.4 ]{Reiprich/Zhang_ea:2011}. 
\item[(iii)] The total cluster mass $M_{500}$ has been determined on the basis of the velocity dispersion of galaxies, 	using spectroscopic data from \citep[tab. 1]{Zhang_ea:2011}. The mass estimator used is eq. (2) of \citep{Biviano_ea:2006}
\beq M_v = A (\frac{\sigma_v}{10^3 \, km s^{-1}})^3 \times 10^{14} h^{-1}\, M_{\astrosun}\, , \label{eq 2 Biviano}
\eeq
$A=1.50 \pm 0.02$, $\sigma_v$ the 3-dimensional velocity dispersion inside a sphere of virial radius (by convention $r_v=r_{200}$).
 Reasons for this choice are given in  \citep[sec. 3]{Biviano_ea:2006}. $M_{500}$ was then determined from $M_{200}$ by a NFW-model.
\end{itemize}

\subsection{\small Theory dependence of  mass data for galaxy clusters\label{subsection theory dependence}}
 Mass densities of the hot gas and of star matter in galaxy clusters are indirectly inferred from observable quantities;  they are  thus  {\em  theory dependent}. Even inside the same background theory they may depend  on  choices of models and methods of evaluation. 

That makes it a difficult task to compare our model with empirical data. A fine-grained judgement  presupposes an evaluation of observational raw data  on the background of WST gravity or, at least, a detailed estimation of systematic errors resulting from a comparison of different background theories (Einstein gravity with $\Lambda CDM$ and Newton approximation, or alternatively TeVeS-MOND, in comparison with WST and  its MOND approximation).  This task 
has to be left to  astronomers, if they  become sufficiently interested in the present approach.   But, taking this caveat in mind,   it still seems possible to confront  available  data from, e.g., the Einstein gravity-$\Lambda CDM$-Newton approximation framework with our model, in order to get a first impression of its potential usefulness.  A  comparison with mass data derived in a TeVeS-MOND background would give  welcome supplementary information. This is not attempted here.

Let us   discuss the possibility and the problems of a confrontation of these data with the MOND-approximation of WST:

\begin{itemize}
	\item[(1)] The mass of the hot gas (intracluster medium) $M_{gas\,500}$ (at $r_{500}$)  has been determined in the mentioned study from X-ray data obtained by  {\em XMM-Newton} and {\em ROSAT}. The temperature of the gas is estimated by a fit to the measured spectrum. The gas density $\rho(r)$ is reconstructed, using model assumptions, from intensity observables and then integrated up to  $r_{500}$.\footnote{In this evaluation the hydrostatic assumption was corrected by taking the velocity dispersion into account \citep[sec. 2.2f.]{Reiprich/Zhang_ea:Corr}. } Up to usual model dependence, the transfer of the mass data from the standard gravity background to WST  seems to be  relatively uncritical.
	
	\item[(2)] Several  methods for determining the stellar mass $M_{\ast \, 500}$ are mentioned  in \citep{Reiprich/Zhang_ea:Corr}. In this study the star mass is gained from optical imaging data due to SDSS 7 in two steps indicated in (ii) above.   According to the authors the difference of the stellar mass estimate due to different initial mass functions can result in a factor 2  \citep[p.4 ]{Reiprich/Zhang_ea:2011}.  Another approach would be to estimate stellar masses of the individual galaxies and ``to construct the stellar mass functions in order to sum the stellar masses'' (ibid, 1). Moreover, an additional component of star matter can be associated to the  intracluster light.  		All in all, the estimate of the star mass concentrated  in galaxies seems to  depend more on models  of galactic star evolution than on the background gravity theory. In spite of that the  precision cannot be expected to be better than by a factor 2 (respectively 0.5).
		
	\item[(3)] In Einstein gravity/$\Lambda CDM$ the cluster  mass can, in principle,  be estimated   from  the velocity dispersion  $\sigma$ of galaxies at distance $R$ (from the center) by an estimator derived from the virial theorem $M \approx G^{-1} \sigma^2 R $. 	The additional acceleration of WST-3L $a_{add}=a_{\varphi}+ a_{sf}$ (\ref{a-add}) is dynamically indistinguishable from the effects of  ``true'' Newtonian masses. So far it seems as if the  estimation of {\em total mass}  can be transferred to the MOND approximation of WST without problems.  
 But if the radius $R$ does not include  the ``whole'' cluster mass (however defined), as is here the case (item (iii) last subsection)  a  surface pressure term must be taken into account. That complicates the case. 

In standard gravity the necessary correction is  implemented by a  cubic mass estimator  $M \sim \sigma^3$ given  above as (\ref{eq 2 Biviano}).  Moroever, the authors of our  reference study \citep{Reiprich/Zhang_ea:2011} reconstruct $M_{500}$ and $r_{500}$ from these values using the NFW profile. Because of the different  profile for the scalar field halo of WST this is a {\em critical} step for our 
exercise.\footnote{An ex-post comparison of the NFW-halo and the WST-halo for the Coma cluster is given in fig. \ref{figure CompNFW}.}  
On the other hand,  if the resulting systematic errors are   smaller than the error intervals of $M_{tot}$ (\ref{total mass}), implied by the observational errors of the other quantities, they do not disturb a rough empirical check of the model.

		\item[(4)] Finally the dependency of the data evaluation on  the background cosmology has to be taken into account. The data of  the 19 clusters used in the following have redshift  $z<0.1$. The geometrical and dynamical corrections implied by the $\Lambda CDM$ cosmology are correspondingly small.  An evaluation in, e.g.,  a Lemaitre-de Sitter model (or even a non-expanding Weyl geometric model with redshift)  \citep[section 4.2]{Scholz:2015MONDlike} would  affect the  data only by a minor expansion of the error intervals.
\end{itemize}
The points (1), (2) and (4), in particular the estimate of stellar mass concentrated in galaxies and gas mass, are fairly insensitive against a change of the background theory from Einstein gravity to WST. The theory dependence of $r_{500}$ is uncritical in our context. Any other reference radius could have been taken, as long as it is   specified  in astronomical distance units. 
 The estimate for the total masses at $r_{500}$ and $r_{200}$   is the critical point  for our purpose (item (3)). However, if the difference  of the  halo profiles between the scalar field energy density of WST and NFW dark matter does not push the estimates for the total masses  $M_{500}, \, M_{200}$ outside the error intervals  of  our halo model (due to observational input data),  we may still be able to draw first inferences from the following evaluation.

\subsection{\small Empirical  data  for 19 clusters \label{subsection observational data}}
\begin{figure}[htbp]
\vspace*{5mm}
    \begin{center}
      \hspace*{-5em}  \includegraphics[scale=0.78,trim=0mm 70mm 5mm 106mm,clip]{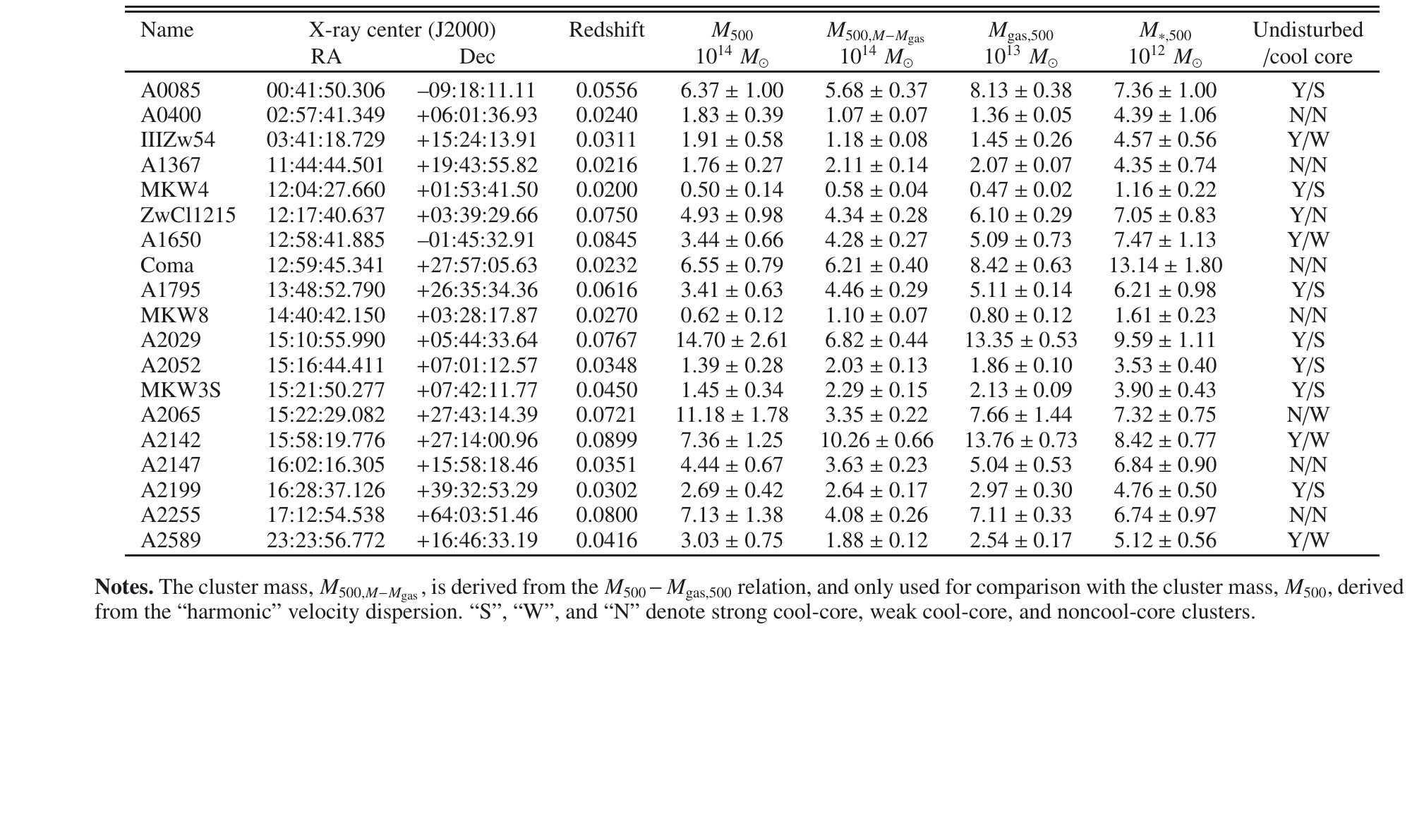}
			\vspace*{-30mm}
    \caption{Tab.1 of \citep{Reiprich/Zhang_ea:Corr}.  Properties of the 19 galaxy clusters}
    \label{Zhang 2012}
    \end{center}
\end{figure}

The studies   \citep{Reiprich/Zhang_ea:2011}, \citep{Reiprich/Zhang_ea:Corr} contain new data on the baryon content and the total gravitational mass for 19 clusters of galaxies  (as it appears in an Einstein--$\Lambda CDM$ framework with Newton approximation).\footnote{\citep{Reiprich/Zhang_ea:Corr} contains a correction to the first paper. Here, of courese, we use the corrected data. }  
 The  mass data $M_{500}, M_{gas\, 500}, M_{\ast\, 500}$ are  given in columns (5), (7), (8) of table 1 in \citep{Reiprich/Zhang_ea:Corr}. It is reproduced in our fig. \ref{Zhang 2012}. The values for $r_{500}$  are  published in   \citep[tab. 1, col. (5)]{Lagana/Reiprich_ea:2011} (here table 1). 
A comparison of the total cluster masses derived from the velocity dispersion with a mass estimate derived from the gas mass shows that the two clusters A2029 and A2065 are outliers, with total cluster masses considerably higher than  the corresponding gas masses would let expect. The authors therefore separate the {\em two outliers} from the rest of the data, with the remaining {\em  17 clusters as a reliable data set} 
\citep[p. 3]{Reiprich/Zhang_ea:2011}. We shall do the same.

Parameters ($\beta, \, r_c$) for  the models of these  galaxy clusters  (as well as of many more) have been published earlier in  \citep[tab. 4.1]{Reiprich:Diss}. This publication also contains mass data  $M_{200},\, M_{gas\, 200}$ at $r_{200}$ (with error intervals),   mass  values $M_A,\, M_{gas\, A}$  (without error interval) at the Abell radius, here defined as  $r_A=2.14\, Mpc$, but no data for star masses.\footnote{Evaluated for the value of $H_o$ assumed in the later publications \citep{Reiprich/Zhang_ea:2011}, \citep{Reiprich/Zhang_ea:Corr}, $h=0.7$.}
 In \citep{Reiprich:Diss}  the methods for determining the total mass and the gas mass were not yet as refined  as in the later  study. It is therefore not possible to aggregate the different data sets to one coherent ensemble.\footnote{The values for $M_{500}$ and $M_{gas500}$ given here differ from the ones in  \citep{Reiprich/Zhang_ea:2011}  outside the error intervals.} 
As no updated values for the parameters ($\beta, \, r_c$)  of the mass profiles  were available to the present author,  the  values of  \citep{Reiprich:Diss} are here used as {\em estimators} for the  form parameters of the $\beta$ model.

We consider $M_{500},\, M_{gas\, 500}, \,M_{\ast\, 500},$ $r_{500}$  from  \citep{Reiprich/Zhang_ea:2011,Reiprich/Zhang_ea:Corr} as  our {\em crucial mass  data}  (including reference radii). Mass data at $r_{200}$ (as well as $r_{200}$ itself) from the older study are welcome  as {\em additional information};  but they will  not be used as core criteria for the empirical test of our model. 
Table 1  collects the data used.\footnote{For error intervals see the respective source, \citep{Reiprich/Zhang_ea:Corr}, \citep{Reiprich:Diss}.} 
It shows that the cluster ensemble covers an order of magnitude variation for the gas mass $M_{gas\, 500}$ and one and a half orders of magnitude variation in total mass $M_{500}$. The selection method by intersecting the cluster sets of different raw data sources does not seem to be influenced by   any particular bias. We thus may consider the collection as a reasonable data set for testing our cluster halo model.

\begin{table}{\center \bf Table 1. Data set used for halo model (error intervals omitted)}\\[0.5ex]
\begin{tabular}{|r||cc|cccc || ccc|} \hline \\[-2.6ex]
{\em Cluster} & $\beta$ & $r_c$ & $r_{500}$ & $M_{500}$ &  $M_{gas\, 500}$ & $M_{\ast\, 500}$ & $M_{200}$ & $r_{200}$ & $M_A$ \\[0.3ex]
\hline \hline
Coma &  0.654 & 246 & 1.278& 6.55& 8.42& 13.14& 13.84& 2.3& 12.86
 \\
 A85 &  0.532 & 59.3 & 1.216 & 6.37 & 8.13 & 7.36 & 7.71 & 1.9 & 8.72\\
A400 & 0.534 & 110. & 0.712 & 1.83 & 1.36 & 4.39 & 1.48 & 1.09 & 2.93\\
IIIZw54 & 0.887 & 206 & 0.731 & 1.91 & 1.45 & 4.57 & 2.81 & 1.35 &  4.51\\
 A1367 & 0.695 & 274 & 0.893 & 1.76 & 2.07 & 4.35 & 4.06 & 1.53 &   5.77\\
 MKW4 & 0.44 & 7.86 & 0.58 & 0.5 & 0.47 & 1.16 & 0.71 & 0.86 &   1.79\\
 ZwCl215 & 0.819 & 308 & 1.098 & 4.93 & 6.1 & 7.05 & 10.37 & 2.09 &   10.65\\
 A1650 & 0.704 & 201 & 1.087 & 3.44 & 5.09 & 7.47 & 11.14 & 2.15 &   11.11\\
 A1795 & 0.596 & 55.7 & 1.118 & 3.41 & 5.11 & 6.21 & 10.99 & 2.14 &   11.04\\
 MKW8 & 0.511 & 76.4 & 0.715 & 0.62 & 0.8 & 1.61 & 2.38 & 1.28 & 4.0\\
 A2029 & 0.582 & 59.3 & 1.275 & 14.7 & 13.35 & 9.59 & 13.42 & 2.29 &   12.59\\
 A2052 & 0.526 & 26.4 & 0.875 & 1.39 & 1.86 & 3.53 & 2.21 & 1.25 &   3.79\\
 MKW3S & 0.581 & 47 & 0.905 & 1.45 & 2.13 & 3.9 & 3.46 & 1.45 &   5.11\\
 A2065 & 1.162 & 493 & 1.008 & 11.18 & 7.66 & 7.32 & 16.69& 2.45 &   14.46\\
 A2142 & 0.591 & 110 & 1.449 & 7.36 & 13.76 & 8.42 & 15.03 & 2.36 &   13.61\\
 A2147 & 0.444 & 170. & 1.064 & 4.44 & 5.04 & 6.84 & 3.46 & 1.45 & 5.15\\
 A2199 & 0.655 & 99.2 & 0.957 & 2.69 & 2.97 & 4.76 & 4.80 & 1.62 &   6.37\\
 A2255 & 0.797 & 423 & 1.072 & 7.13 & 7.11 & 6.74 & 13.32 & 2.27&   12.52\\
 A2589 & 0.596 & 84.3 & 0.848 & 3.03 & 2.54 & 5.12 & 3.58 & 1.47 &   5.24\\
\hline
\end{tabular}\\[0.5ex]
{\em Source:} first and last block \citep{Reiprich:Diss}, middle block \citep{Reiprich/Zhang_ea:Corr}, $r_{500}$ \citep{Lagana/Reiprich_ea:2011} \\
{\em Units:}  $r_c$ in $kpc$, $r_{500}, r_{200}$ in $Mpc$, $M_{500}$ in $10^{14}$ $M_{\astrosun}$, $ M_{gas\, 500}$ in $10^{13}$ $M_{\astrosun}$, $ M_{\ast\, 500}$ in $10^{12}$ $M_{\astrosun}$,
\end{table}

\subsection{\small Adapting the WST halo model  to the data  \label{WST data}}

For the construction of the model we have to work in  Einstein gauge (\ref{Einstein gauge}). Consistency with the deep MOND acceleration demands (\ref{a-tilde}). Taken together these conditions fix the coefficients of the Lagrangian (\ref{modified Hilbert term}) and (\ref{cubic Lagrangian}),  independently of the convention chosen for $\phi_o$. 
With a value for  $\lambda$  on the order of magnitude 10 (e.g., $\frac{\lambda}{4}=6$ like at the end of section \ref{subsection WST-3L}) the contribution of the $L_{V4}$ term  lies many orders of magnitude below the energy densities the dominant term in  (\ref{ThetaI}) and is negligible in our context. 

In agreement with section \ref{section halos}   the test of our halo model for each of the 19 galaxy clusters can now proceed as follows: 
\begin{itemize}
	\item[(1)] Specification  of the $\beta$ models (\ref{beta model}) for gas and for star mass;  parameters ($\beta,\, r_c$)  from \citep{Reiprich:Diss}, $\rho_o$ determined by fitting to $M_{gas\,500}$ and $M_{\ast\,500}$ at $r_{500}$   respectively \citep{Reiprich/Zhang_ea:Corr}.\footnote{If one wants to investigate the external halo (beyond $r_{200}$) one has to choose a  cutoff at $r_{200}$ or fading out functions. The main part of of our investigation deals with the internal halo; only in the final discussion, section \ref{section discussion},  questions of the external halo come into the play. This remark applies, {\em mutatis mutandis}, to items (3), (4) below \label{fn fade out} }
	\item[(2)] Determination of the Newton accelerations of the baryonic mass components.
	\item[(3)] Calculation of the scalar field halo and the phantom halo of the baryonic mass by (\ref{scalar field halo baryonic matter}),  (\ref{phantom halo}).
	\item[(4)] Calculation of the scalar field halo of the system of freely falling galaxies  (\ref{scalar field halo galaxies}).
	\item[(5)] Aggregation of these to the total halo (\ref{transparent matter halo});  choice of fading out functions beyond $r_{200}$ (see appendix).
	\item[(6)] Integration of the densitities to the  corresponding masses: 
 scalar field energy of the galaxies $M_{s\hspace{-0.1em}f \, 2}$, of the gas mass $M_{s\hspace{-0.1em}f \, 1}$, total scalar field halo $M_{s\hspace{-0.1em}f }$, net phantom energy of the baryonic mass $M_{ph\,1}$, finally the total transparent matter   $M_{t}$  and the lensing mass $M_{lens}$ (\ref{total mass}), (\ref{lensing mass}).
	\item[(7)] Calculation of the error intervals of the model at selected distances ($r_{500},\,  r_{200}$).
	\item[(8)] Comparison of the empirical value for $M_{500}$ (respectively  $M_{200}$) with the model value $M_{t}(r_{500})$ (resp.  $M_{t}(r_{200})$).
\end{itemize}
The result of this test are  given in subsection \ref{subsection 19 clusters}. Before we turn to the overall evaluation, we shall have a look at one cluster as an exemplary case.

\subsection{\small The WST halo model with the Coma cluster as  test case \label{subsection Coma}}
\begin{figure}
\center{\hspace*{0em} \includegraphics*[scale=1.2]{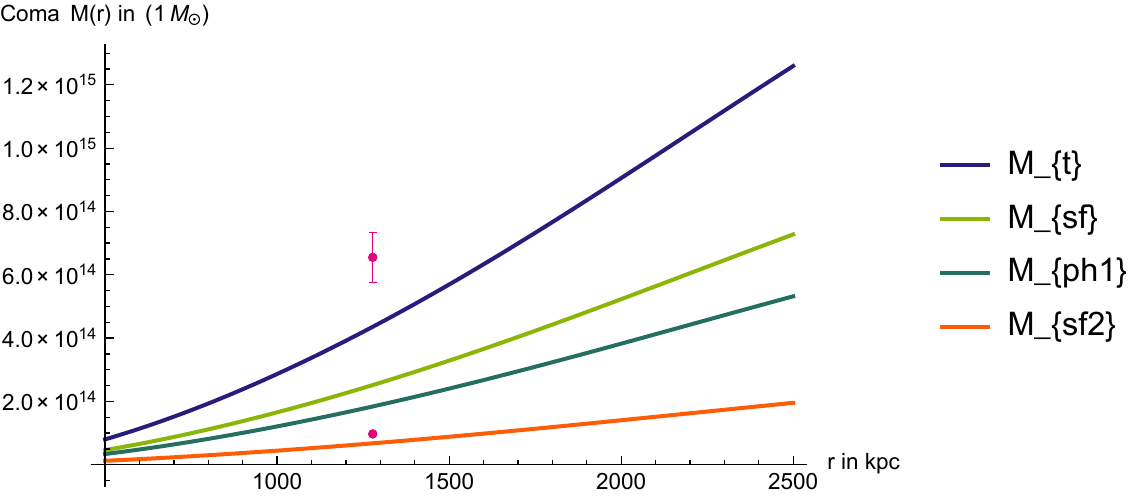}}
\caption{\em  Halo components of Coma cluster:  transparent matter halo
 $M_{t}= M_{s\hspace{-0.1em}f }+M_{ph\,1}$,  total scalar field  (SF) halo 
$M_{s\hspace{-0.1em}f }$,   halo of freely falling galaxies  $M_{s\hspace{-0.1em}f \, 2}$, and 
  net phantom energy  $M_{ph\,1}$ (in barycentric rest system). \newline 
Empirical data (violet dot and bar):  baryonic mass $M_{\ast\,500}+M_{gas\,500}$ (dot) and $M_{500}$ (with error intervals) at $r_{500}= 1280 \, kpc$. }
\label{fig halo Coma}
\end{figure}

For a first check of our model we choose the {\em Coma cluster}.
 According to item (1) of the last section we use the  empirical input data from table 1 (units given there):
\beq (\beta, r_c,  r_{500},M_{gas\,500},M_{\ast\,500}) =(0.654, \, 246, \, 1.278, \, 8.42, \, 13.14  )\, , \label{model parameters Coma}
\eeq 
($r_c$ in $kpc$, $r_{500}$ in $Mpc$, $M_{gas\,500}$ in $10^{13} M_{\astrosun}$, $M_{\ast\,500}$ in $10^{12} M_{\astrosun}$).
The different components of the cluster halo integrate to  masses (up to distance $r$) documented in fig. \ref{fig halo Coma}.

\begin{figure}
\center{\hspace*{0em} \includegraphics*[scale=1.2]{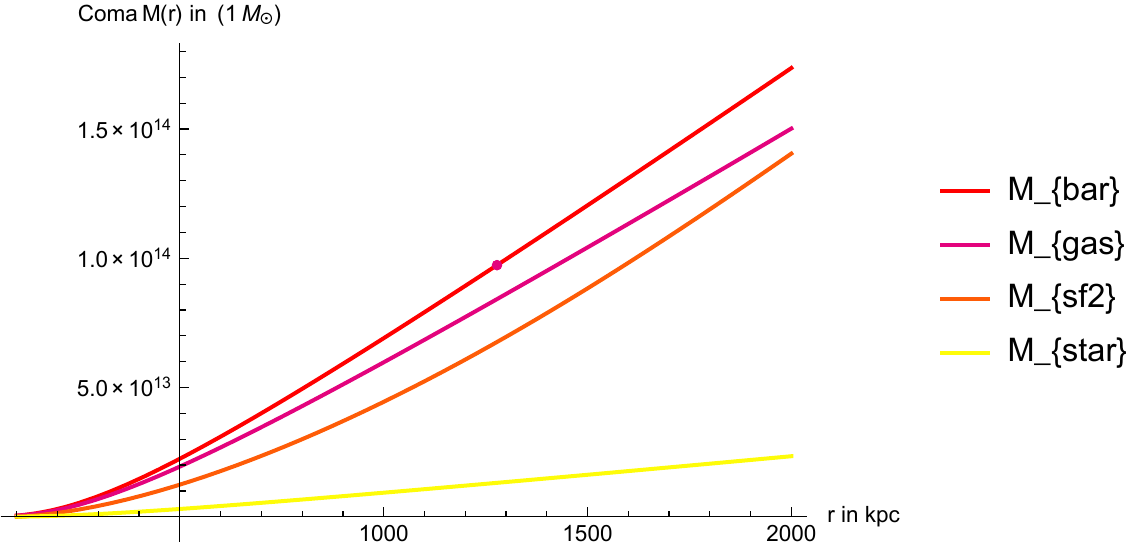}}
\caption{\em  Comparison of the contribution of the scalar field halo of the  galaxies $M_{sf2}$ with the baryonic mass for the Coma cluster (empirical data for $ M_{bar\,500}$ violet dot).}
\label{SF star comparative}
\end{figure}

The scalar field halo  (SF)  of the total baryonic mass    $\rho_{s\hspace{-0.1em}f  \, 1}$ contributes the lion share to the total transparent energy. The SF of the galaxy system $\rho_{s\hspace{-0.1em}f \, 2}$ carries about as much energy as the net phantom component $\rho_{ph\,1}$ in the barycentric rest system of the cluster.  It   surpasses the gas mass close to  the reference radius  $r_{500}$. 
 Figure \ref{SF star comparative} shows the fast increase of the gravitational mass of the scalar field halo of the galactic system.

\begin{figure}
\center{\includegraphics*[scale=1.2]{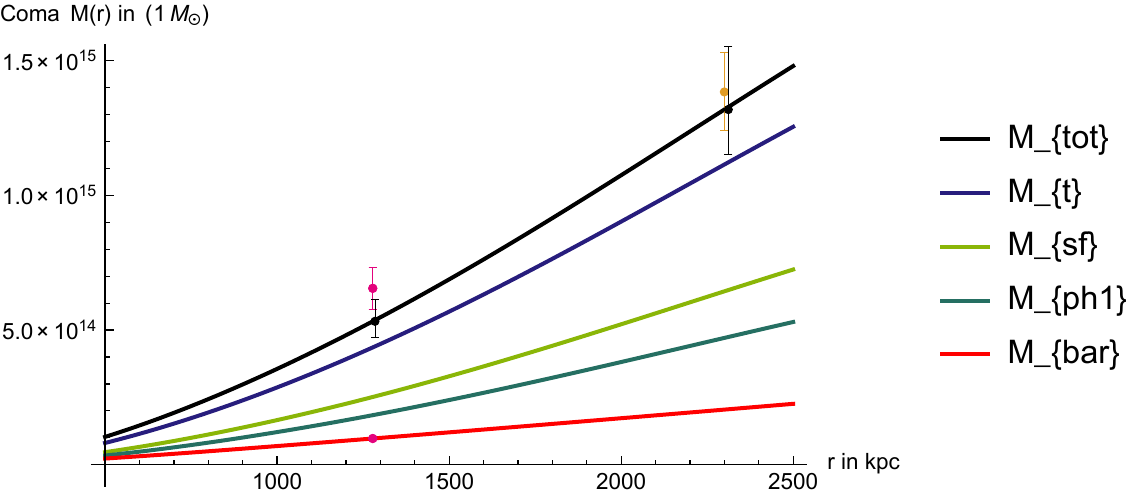}}
\caption{\em Contribution of the baryonic mass $M_{bar}$,  of the scalar field and the phantom energies $M_{s\hspace{-0.1em}f }, \, M_{ph\,1}$  to the transparent mass $M_{t}$ and to the  total mass   $M_{tot}=M_t+M_{bar}$ of the Coma cluster   in the WST model.  Model errors indicated at $r_{500}$,  $ r_{200}$ (black). Empirical data for $M_{bar}$ (violet dot)  and for the empirically determined total mass $M_{500}$ with error bars (violet)  from \citep{Reiprich/Zhang_ea:Corr}. Additional empirical data  at $ r_{200}$ (yellowish) from  \citep{Reiprich:Diss}. }
\label{fig Coma}
\end{figure}

If we add all baryonic and halo contributions, the picture given in fig. \ref{fig Coma} emerges. 
It shows an encouraging agreement of the observed values  $M_{500}$ at $r_{500}= 1278\, kpc$  with the prediction of the Weyl geometric  halo model $M_{tot}(r_{500})$   (in the range of the observational errors and of model errors). 
\beq M_{tot}(r_{500}) =5.66^{+0.97 \atop -0.68}\,  , \qquad M_{500}= 6.55^{+0.79 \atop -0.79} \quad   \times 10^{14}\, M_{\astrosun}
\eeq 
Model error bars have been estimated by varying the  input data (\ref{model parameters Coma}) in their respective error intervals:\footnote{The reconstruction of star mass from observational raw data is a particularly delicate point. Depending on the assumptions on the stellar dynamics and the resulting data evaluation model ``one can obtain up to a factor of 2 fewer stars'' \citep[p. 6]{Reiprich/Zhang_ea:Corr}. For obtaining our model errors we allowed a variation in stellar masses by factors 0.5 and 2.} 

At the Abell radius $r_A=2.14\, Mpc$ (in the convention of \citep{Reiprich:Diss}) we get the model values
\beq M_{tot}(r_A) \approx 11.9\, , \quad M_{lens}(r_A) \approx 7.7 \quad \times 1 0^{14}\,M_{\astrosun} 
\eeq 
 for the dynamical total mass and the lensing mass (\ref{lensing mass}).\footnote{The total (dynamical) transparent matter contribution is $M_{t}(r_A) = 9.30$, of which $M_{s\hspace{-0.1em}f \, 2}(r_A) \approx   2.01$ are due to the scalar field halo of the galaxy system. The  net phantom energy  amounts to  $M_{ph\,1}(r_A)\approx  1.82  $ (all values in units of $  10^{14}M_{\astrosun}$).} 
Due to the net phantom energy  the dynamical mass is about 20 \% higher than the lensing mass. This is an effect by which the model can, in principle, be  tested empirically and discriminated from others.\footnote{The lensing data of \citep{Gavazzi_ea:2009}  seem consistent with such an observation, $M_{lens}(r_1)=6.1^{+12.1 \atop -3.5} \times10^{14}\, M_{\astrosun}$ at $r_1=2.5^{+0.8 \atop -0.5}\, Mpc$, compared with $M_A\approx 12.9\times 10^{14}\, M_{\astrosun}$ at $r_A \approx 2.14\, Mpc$ in \citep{Reiprich:Diss}.   But because of their exorbitant large error intervals these data  are far from significant for our question.}
The baryonic masses in the $\beta$ model are $M_{gas}(r_A) \approx 11.73\times   10^{13}\,M_{\astrosun}$, $M_{\ast}(r_A) \approx 1.83 \times 10^{13}\,M_{\astrosun}$. 
Thus not only the  total mass $M_{tot}(r_{500})$ given by our model agrees with the  empirical value  $M_{500}$ inside the error bounds;  also $M_A$ and $M_{200}$ are   reasonably well recovered. 

This is the case {\em without  assuming}  any  component  {\em of particle   dark matter}  besides the (real) energy of the scalar field and the (phantom) energy ascribed to the additional acceleration $a_{\varphi}$ induced by the   Weylian scale connection. This might still be  a coincidence. In order to learn more about the question whether the findings at the Coma cluster  are exemplary or not, we have to consider the  data of all 19  galaxy clusters, respectively  the 17 of the reliable sub-ensemble. 

\subsection{\small Halos and total mass for 17(+2) clusters of galaxies \label{subsection 19 clusters}}
\begin{figure}[htb]
	\hspace*{-5em}			\includegraphics*[scale=0.65]{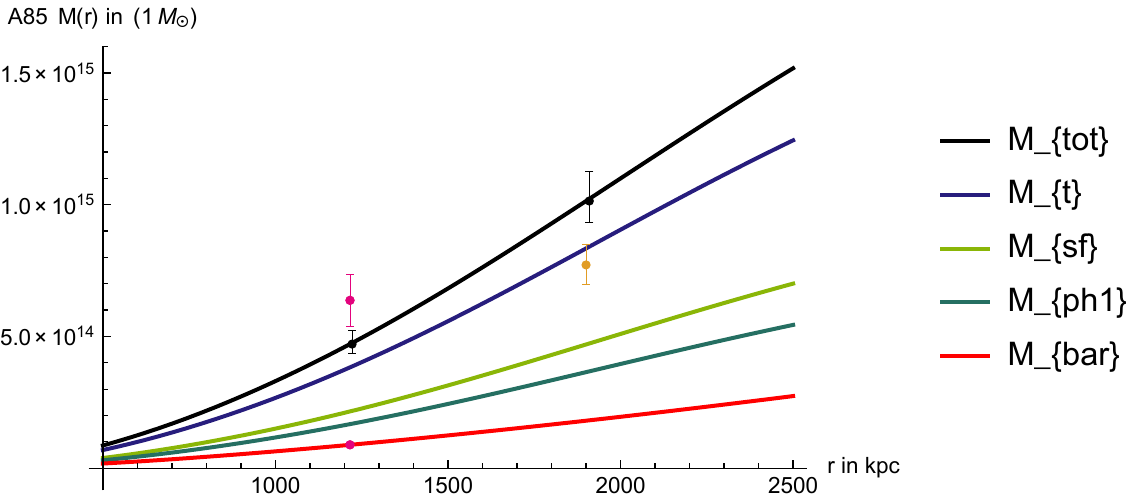}
	\qquad	\includegraphics*[scale=0.65]{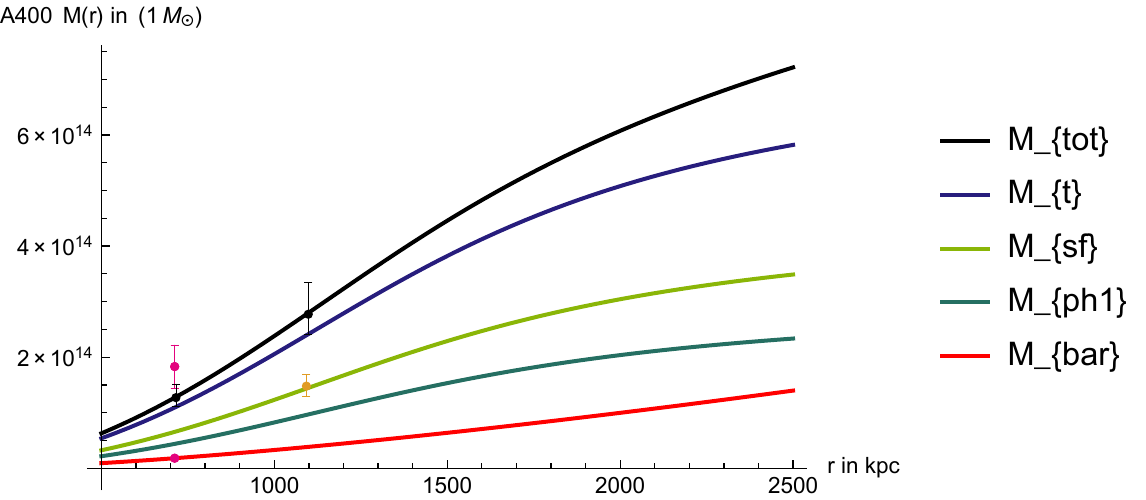}
		\hspace*{-5em}				\includegraphics*[scale=0.65]{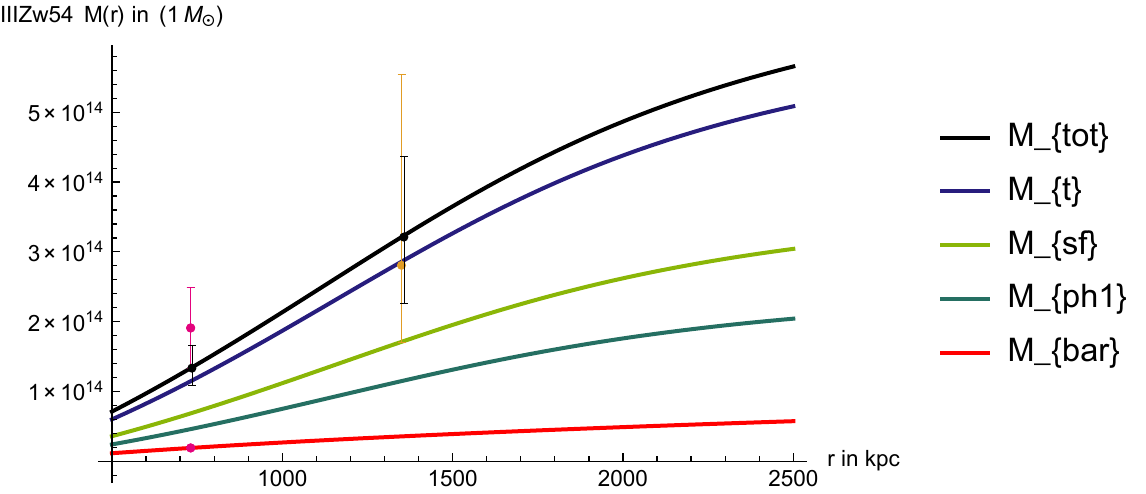}
		\qquad	\includegraphics*[scale=0.65]{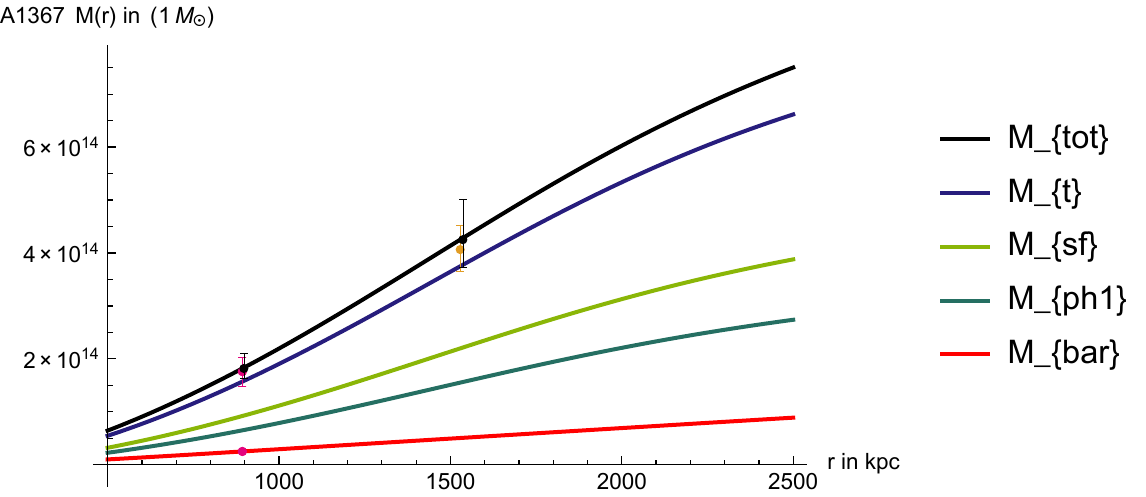}
			\hspace*{-5em}					\includegraphics*[scale=0.65]{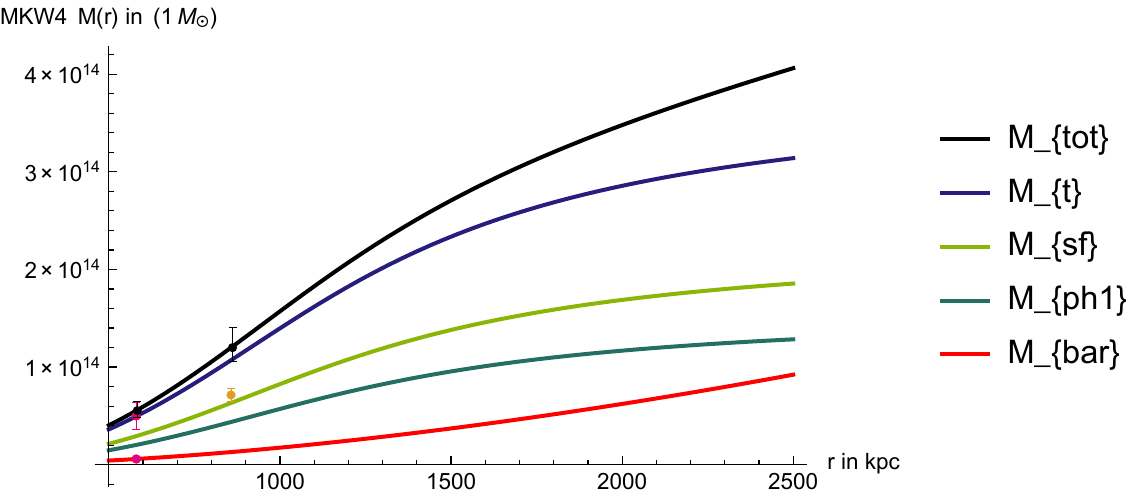}\qquad	\includegraphics*[scale=0.65]{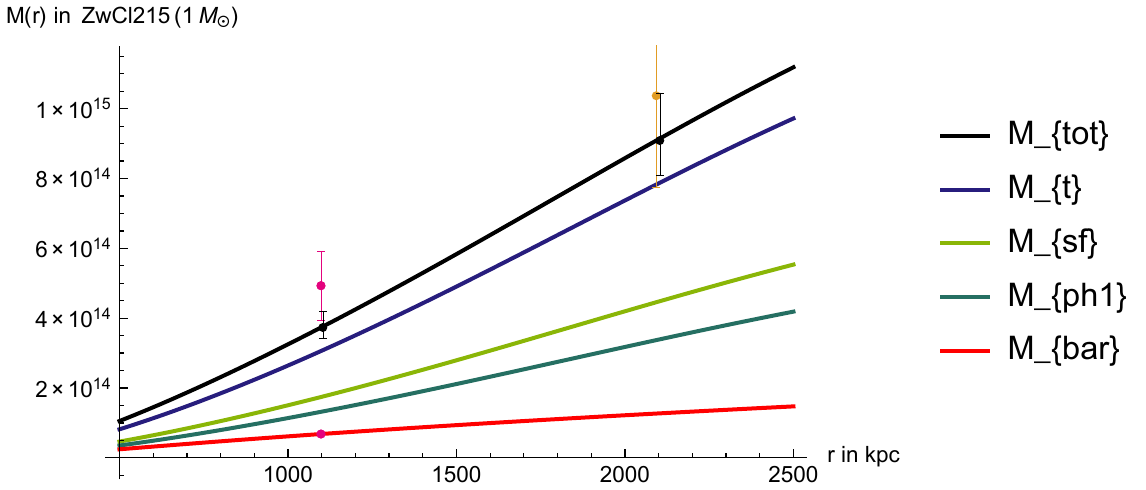}
				\hspace*{-5em}						\includegraphics*[scale=0.65]{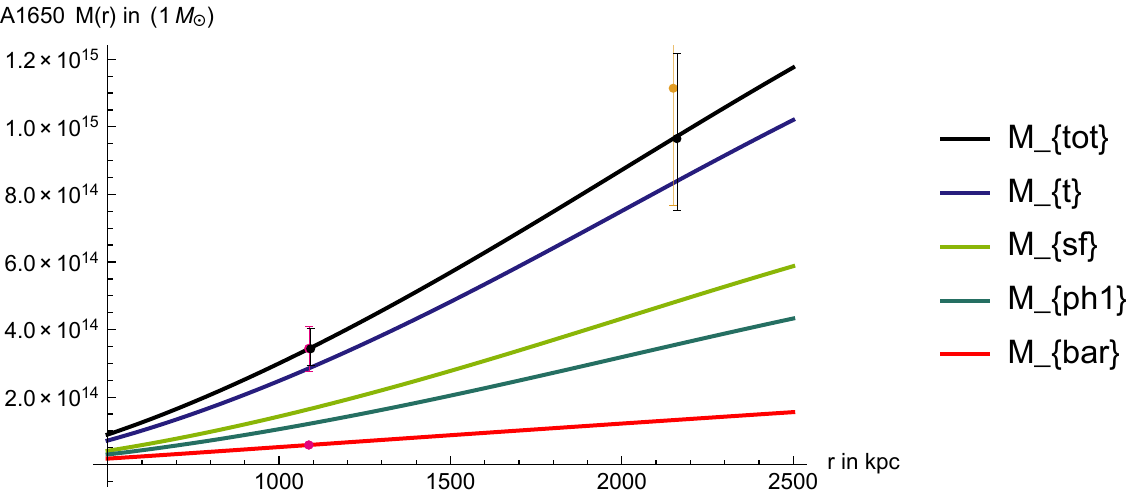}\qquad	\includegraphics*[scale=0.65]{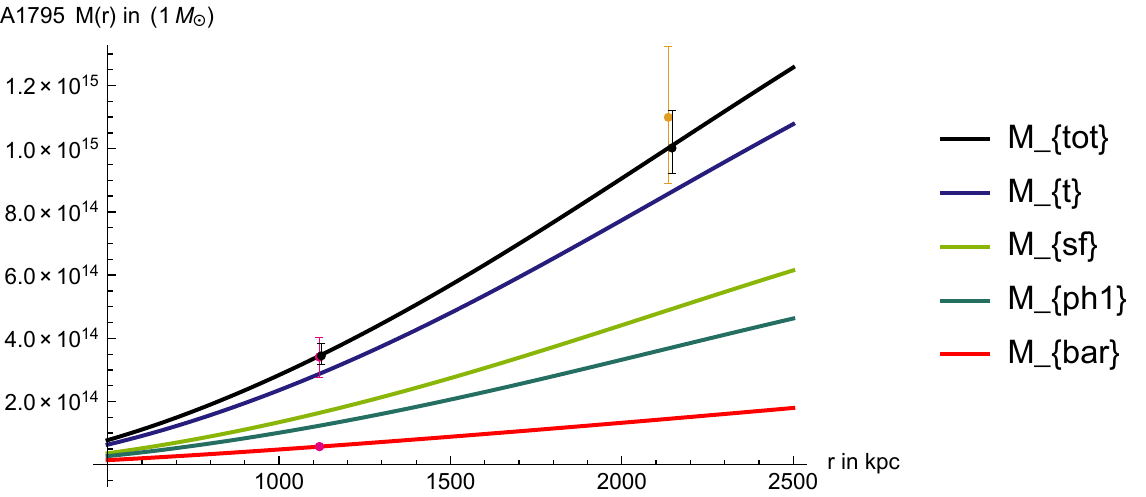}
				
							\captionof{figure}{\em Halo models  for clusters 2 -- 9 in tab. 1.:  total mass  $M_{tot}$ (black line) with model error bars at $r_{500},\, r_{200}$, transparent matter halo $M_{t}$ constituted by scalar field halo $M_{s\hspace{-0.1em}f \, 2}$ and net phantom halo (in barycentric rest system of cluster) $M_{ph\,1}$ and baryonic mass (gas and stars) $M_{bar}$. 
							Empirical data  for the total mass with error intervals at $r_{500}$ (violet)  from \citep{Reiprich/Zhang_ea:Corr}. Additional empirical data  at $r_{200}$ (yellow) from  \citep{Reiprich:Diss}. }
								\label{fig clustersI}
\end{figure}

\begin{figure*}[htb]
		\hspace*{-5em}\includegraphics*[scale=0.65]{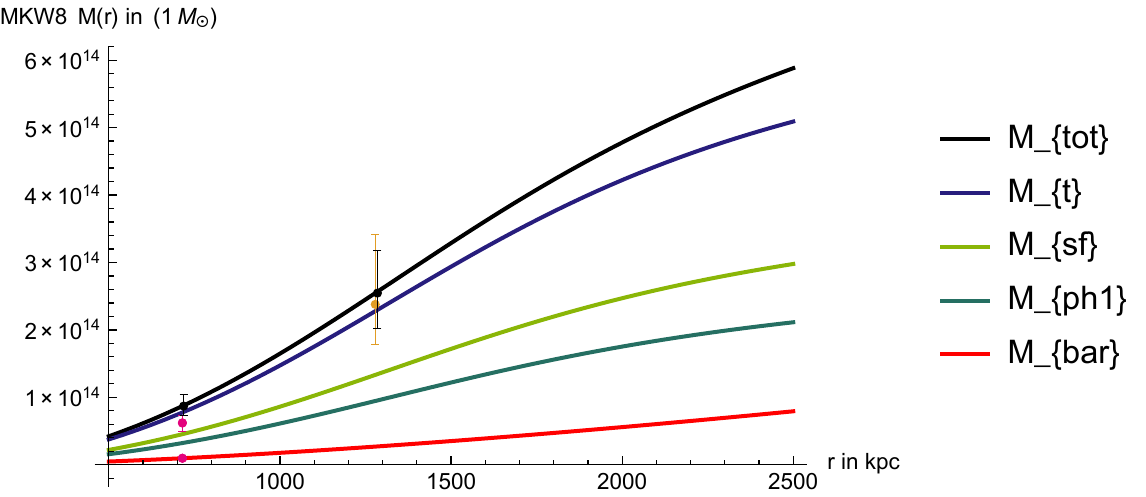}\qquad	
		\includegraphics*[scale=0.65]{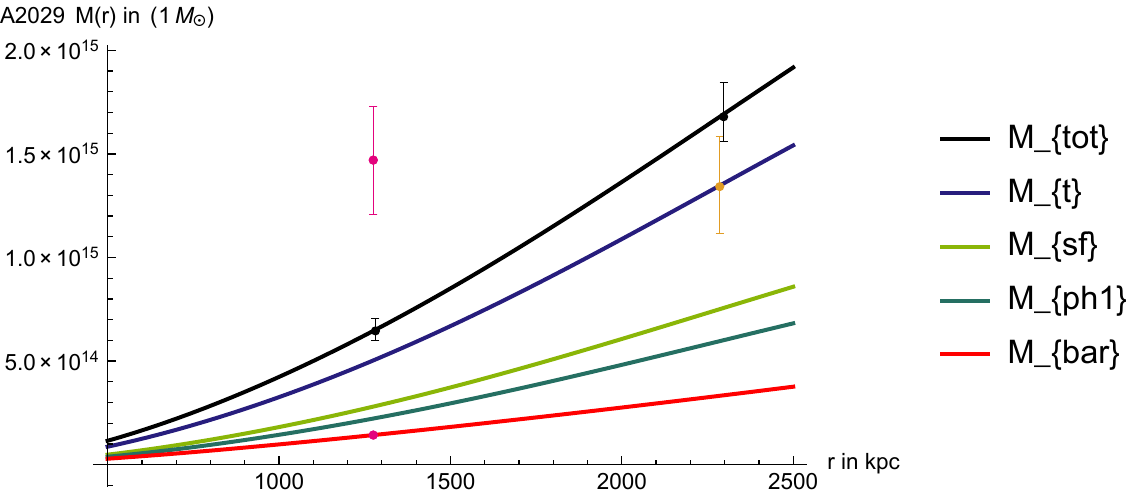}
					\hspace*{-5em}	\includegraphics*[scale=0.65]{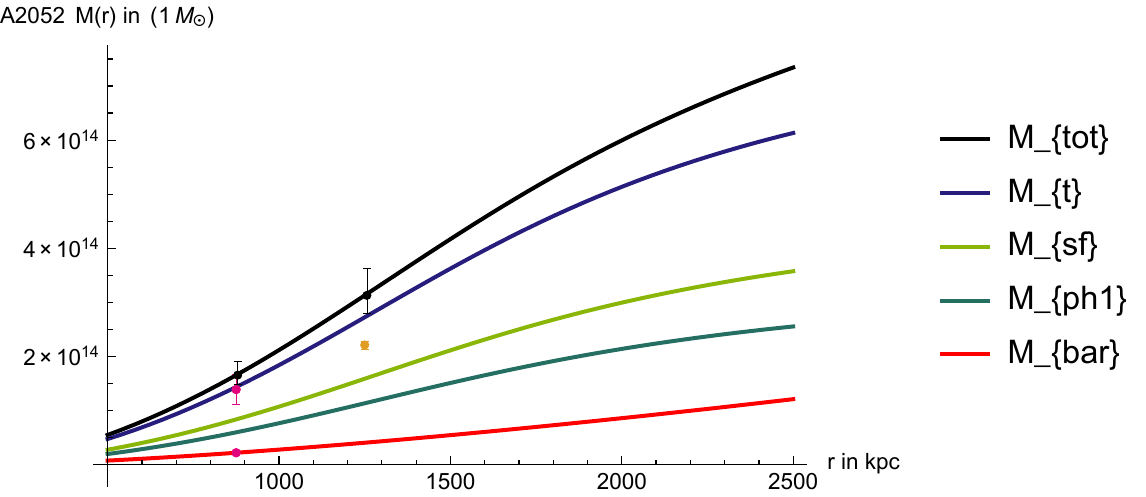}\qquad	
				\includegraphics*[scale=0.65]{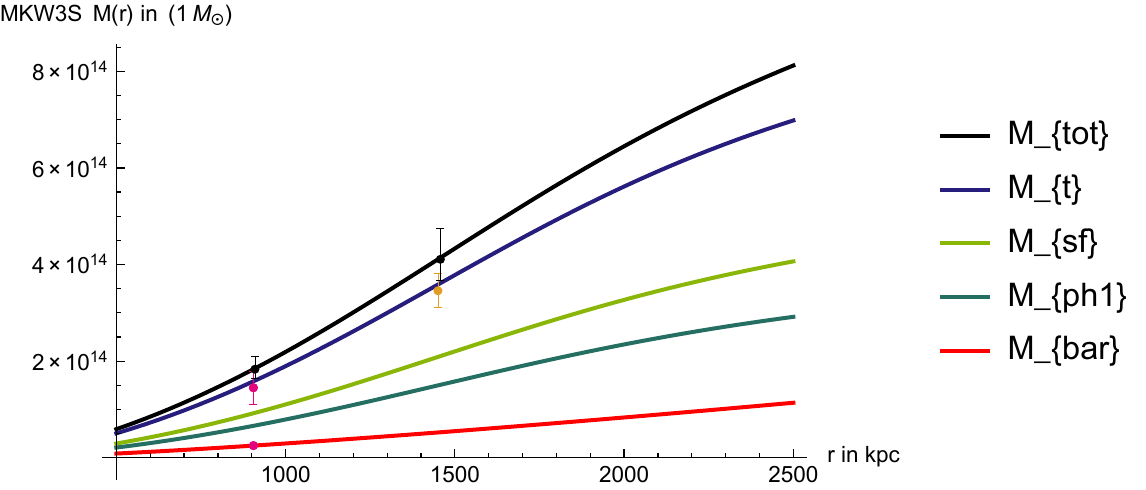}
						\hspace*{-5em}		\includegraphics*[scale=0.65]{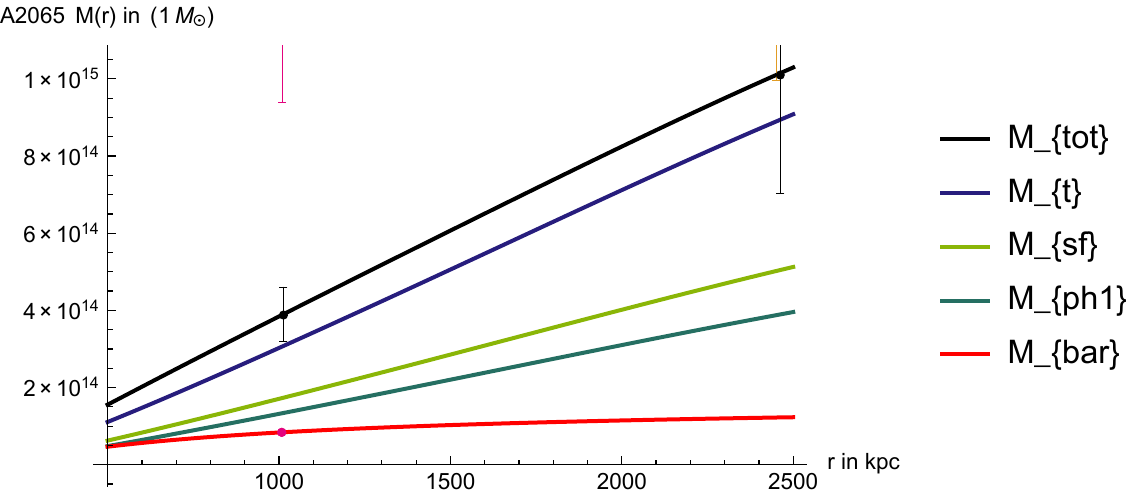}\qquad	\includegraphics*[scale=0.65]{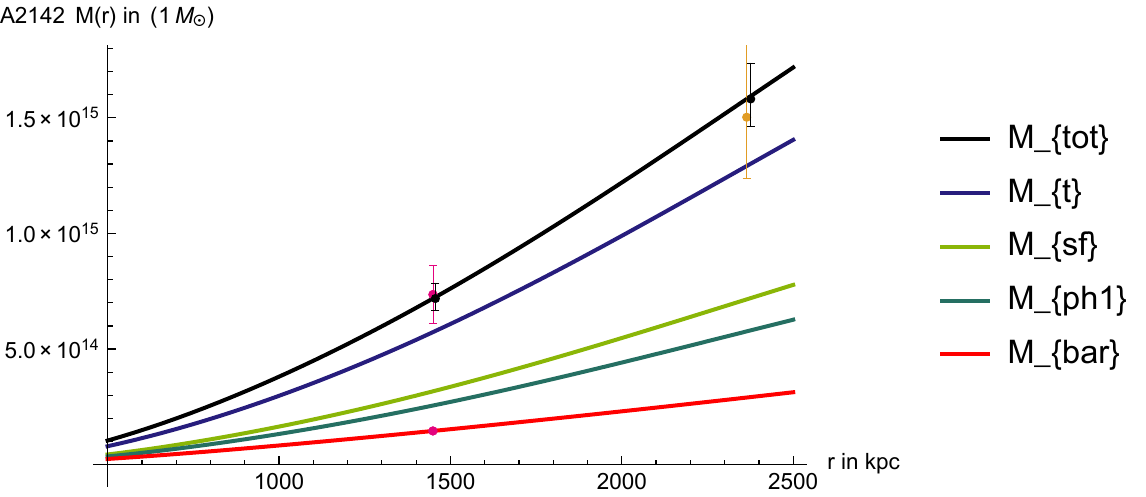}
					\hspace*{-5em}					\includegraphics*[scale=0.65]{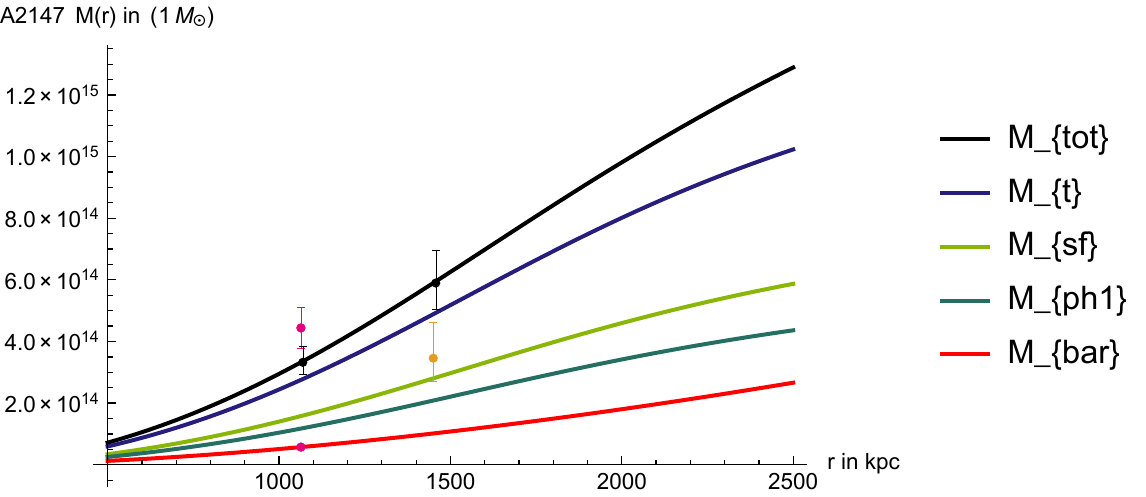}\qquad 	\includegraphics*[scale=0.65]{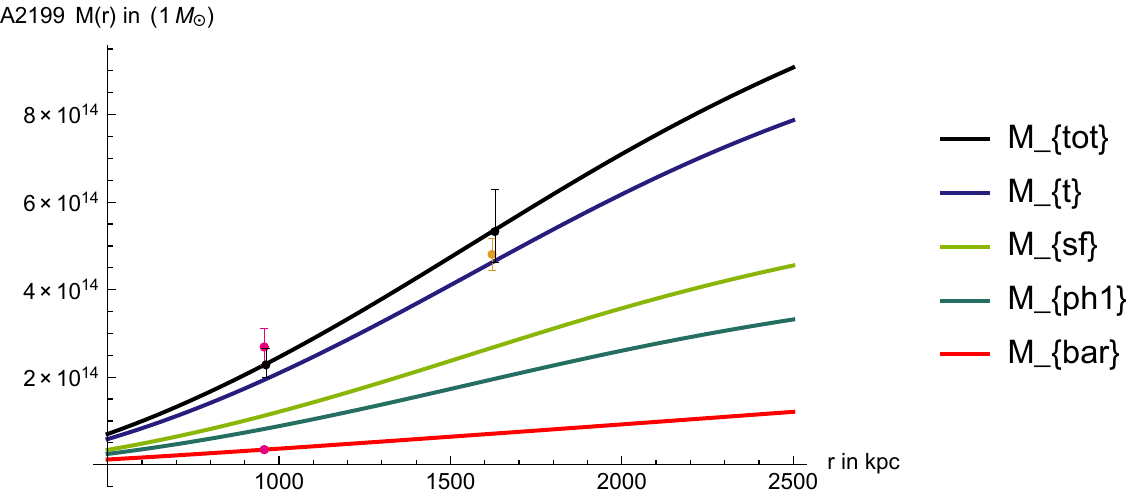}
								\hspace*{-5em}								\includegraphics*[scale=0.65]{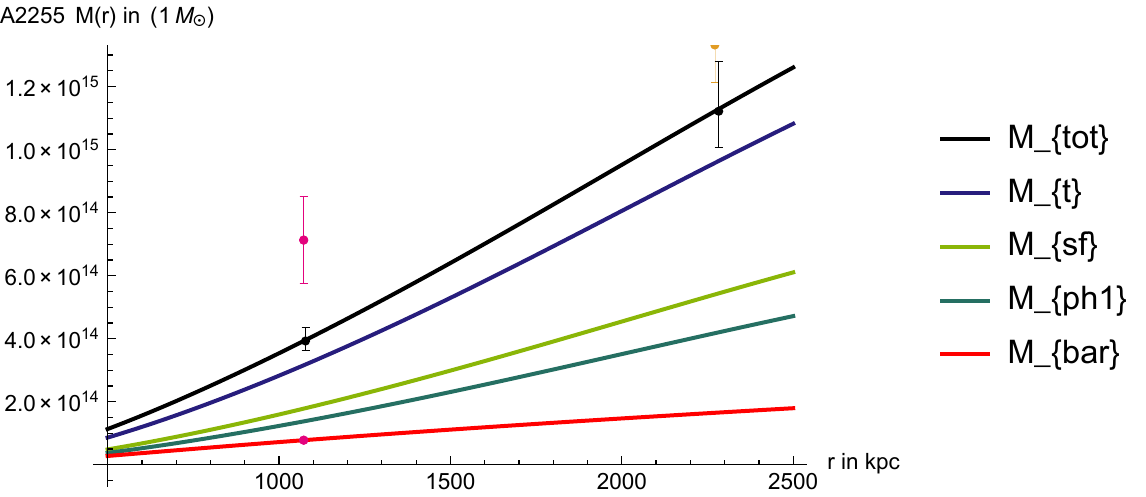} \qquad	\includegraphics*[scale=0.65]{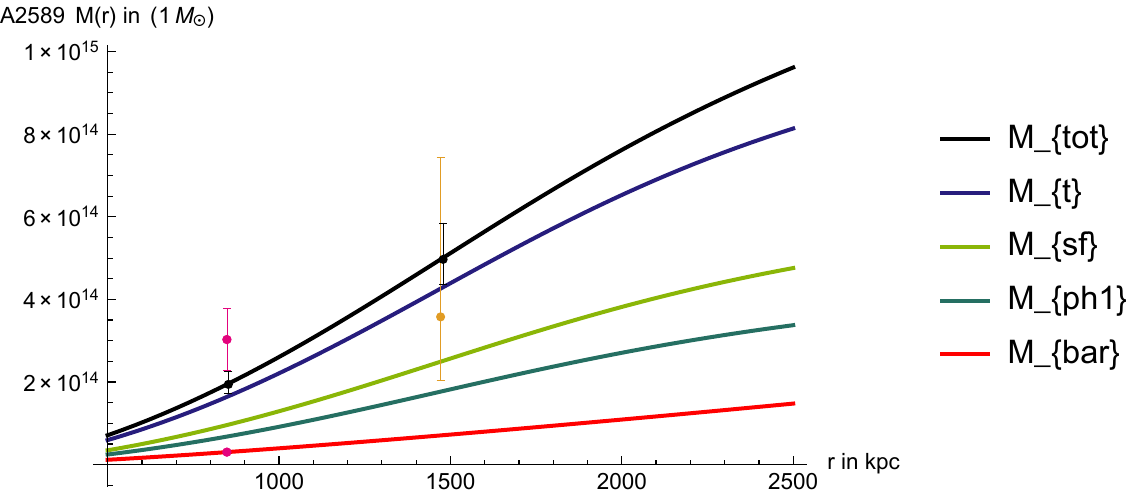}
							\caption{\em Halo models  for clusters 10 -- 19 in tab. 1. Description  see fig. 	\ref{fig clustersI}}
							\label{fig clustersII}
\end{figure*}

The mass values of the halo models for the 19 clusters are calculated as described in section \ref{subsection observational data} with the choice of fadeout functions beyond $r=r_{200}$ (see appendix). The {\em results}  are documented in tables 2, 3 and figs. \ref{fig clustersI}, \ref{fig clustersII} (for Coma see fig. 4). The baryonic masse at $r_{200}$ ($M_{gas}(r_{200}), M_{\ast}(r_{200})$) of table 3 result from an extrapolation of the empirical values at $r_{500}$ given in \citep{Reiprich/Zhang_ea:Corr} to $r_{200}$ using the $\beta$-models of section 3.4, item (1), for gas and star mass.

 {  For 15 clusters} our {\em model reproduces}  the total mass at $r_{500}$  {\em correctly, i.e. inside the error margins of data and model}. This is achieved without any further adjustable parameter, only on  the basis of the parameters for the $\beta$-model for baryonic mass and $M_{gas}(r_{500}), M_{\ast}(r_{500})$. The fading out functions do not  intervene below $r_{200}$.  Moreover, for the majority  of these,  and paradoxically  for all other four,  the less precisely determined data at $r_{200}$ have overlapping $1 \sigma$ error intervals. This indicates a surprising agreement between the (theoretically derived) transparent halo and the empirically determined dark halo, $M_t \approx M_{dm}$.

{\em For 5 clusters}, A85, A2255, A2589 and the  outliers A2029, A2065, the  { error intervals} of empirical  data and model data  {\em  do not overlap}.  For the first three of them (A85, A2255, A2589)  the model predictions are consistent with the empirical data  within  doubled error intervals ($2\sigma$ range).  Only the  two outliers (A2029, A2065)  lie farther apart.\footnote{A2029 has the surprising property that the  empirical values for the {  total mass at} $r_{500}$  {\em surpasses}  the one at $r_{200}$, $M_{500}> M_{200}$.} 
Otherwise the model data are  in good  agreement with an assumption of normally distributed statistical  errors and with the assumption that the {\em evaluation bias} due to the use of the   NFW profile for dark matter (item (3) in section \ref{subsection theory dependence}) {\em does not shift the mass estimates outside the error intervals}. 

 All in all,  the { assessment of  the WST-3L halo model}  
has surprisingly  well  {  passed}, in spite of  the  main caveat of item (3), section \ref{subsection theory dependence}.  The outcome  found for    Coma seems to be typical also for the other clusters.  
Moreover, the good agreement of the model with the data of the reliable  sub-ensemble of 17 clusters   supports the assumption stated in the last phrase  (3), section \ref{subsection theory dependence} (no large systematic errors due to data transfer from Einstein/$\Lambda CDM$ to the WST-3L framework). But still we cannot exclude the possibility of cancelling between model errors and data transfer errors.  Thus we have only found empirical support for the {\em conjecture} that, on the level of galaxy clusters, the observed dark matter effects encoded in $M_{dm}$  may solely be due  to the combined impact of the  halo $M_{sf}$ of the scalar field and of the scale connection, $M_{ph\,1}$ (cf. (\ref{total mass})):
\beq M_{dm} \approx  M_{sf} +  M_{ph\,1} =  M_t \qquad  \label{conjecture}
\eeq

\begin{table}{\center \bf Table 2. Empirical values ($M_{500},  M_{200}$) and model values ($M_{tot}$) for total mass at $r_{500}, \, r_{200}$}\\[0.5ex]
\begin{tabular}{|r||rrr|rrr|} \hline \\[-2.6ex]
{\em Cluster}& $r_{500}$& $M_{tot}(r_{500})$& $M_{500}$&$r_{200}$& 
      $M_{tot}(r_{200})$& $M_{200}$ \quad\\ \hline  \hline \\[-2.2ex]
	Coma & 1278. & 5.32 $^{+ 0.83 \atop 
      -0.58 }$ & 6.55 $^{+ 0.79 \atop -0.79 }$ & 2300. 
      & 13.19 $^{+ 2.36 \atop -1.65 }$ & 
      13.84 $^{+ 1.49 \atop -1.41 }$\\ 
A85 & 1216. & 4.67 
      $^{+ 0.51 \atop -0.36 }$ & 6.37 $^{+ 1.00 \atop -1.00 
      }$ & 1900. & 9.70 $^{+ 1.09 \atop 
      -0.77 }$ & 7.71 $^{+ 0.8 \atop -0.74 }$\\ 
A400 & 712. & 1.26 $^{+ 0.25 \atop -0.16 
      }$ & 1.83 $^{+ 0.39 \atop -0.39 }$ & 
      1093. & 2.62 $^{+ 0.55 \atop -0.35 }$ & 
      1.48 $^{+ 0.21 \atop -0.18 }$\\ 
IIIZw54 & 731. & 
      1.33 $^{+ 0.34 \atop -0.27 }$ & 1.91 
      $^{+ 0.58 \atop -0.58 }$ & 1350. & 3.09 $^{+ 1.30 
      \atop -1.02 }$ & 2.81 $^{+ 2.74 \atop -1.10 }$\\ 
A1367 & 893. & 1.82 $^{+ 0.29 \atop -0.19 }$ & 
      1.76 $^{+ 0.27 \atop -0.54 }$ & 1529. & 
      4.28 $^{+ 0.81 \atop -0.62 }$ & 4.06 
      $^{+ 0.45 \atop -0.40 }$\\ 
MKW4 & 580. & 0.54 $^{+ 
      0.09 \atop -0.06 }$ & 0.50 $^{+ 0.14 \atop -0.14 }$ 
      & 857. & 1.10 $^{+ 0.19 \atop -0.13 }$ 
      & 0.71 $^{+ 0.07 \atop -0.06 }$\\ 
ZwCl215 & 1098. & 3.74 $^{+ 0.47 \atop -0.32 }$ & 
      4.93 $^{+ 0.98 \atop -0.98 }$ & 2093. & 
      9.32 $^{+ 1.49 \atop -1.07 }$ & 
      10.37 $^{+ 3.51 \atop -2.62 }$\\ 
 A1650 & 1087. & 3.42 $^{+ 0.63 \atop -0.50 }$ & 
      3.44 $^{+ 0.66 \atop -0.66 }$ & 2150. & 9.40 
      $^{+ 2.87 \atop -2.31 }$ & 11.14 $^{+ 
      5.77 \atop -3.46 }$\\
 A1795 & 1118. & 
      3.40 $^{+ 0.40 \atop -0.26 }$ & 3.41 $^{+ 0.63 
      \atop -0.63 }$ & 2136. & 9.32 $^{+ 1.11 
      \atop -0.74 }$ & 10.99 $^{+ 2.26 \atop 
      -2.09 }$\\ 
MKW8 & 715. & 0.86 $^{+ 0.18 
      \atop -0.14 }$ & 0.62 $^{+ 0.12 \atop -0.12 }$ & 
      1279. & 2.35 $^{+ 0.67 \atop 
      -0.54 }$ & 2.38 $^{+ 1.04 \atop -0.59 }$\\ 
 A2029 & 1275. & 6.39 $^{+ 0.61 \atop -0.43 }$ & 
      14.7 $^{+ 2.61 \atop -2.61 }$ & 
      2286. & 15.83 $^{+ 1.57 
      \atop -1.12 }$ & 13.42 $^{+ 2.43 \atop 
      -2.26 }$\\ 
A2052 & 875. & 1.63 $^{+ 0.26 
      \atop -0.17 }$ & 1.39 $^{+ 0.28 \atop 
      -0.28 }$ & 1250. & 2.94 $^{+ 0.47 \atop 
      -0.32 }$ & 2.21 $^{+ 0.06 \atop -0.08 
      }$\\ 
MKW3S & 905. & 1.90 $^{+ 0.27 \atop -0.18 }$ 
      & 1.45 $^{+ 0.34 \atop -0.34 }$ & 1450
      & 3.81 $^{+ 0.60 \atop -0.40 }$ & 3.46 $^{+ 0.36 
      \atop -0.34 }$\\ 
A2065 & 1008. & 3.92 $^{+ 0.72 
      \atop -0.66 }$ & 11.18 $^{+ 1.78 \atop -1.78 }$ 
      & 2450. & 11.90 $^{+ 8.20 \atop -4.61 
      }$ & 16.69 $^{+ 21.34 \atop -6.73 }$\\ 
A2142 & 1449. & 7.15 $^{+ 0.68 \atop -0.50 }$ & 
      7.36 $^{+ 1.25 \atop -1.25 }$ & 2364. & 
      15.29 $^{+ 1.53 \atop 
      -1.14 }$ & 15.03 $^{+ 3.9 
      \atop -2.64 }$\\ 
A2147 & 1064. & 3.32 
      $^{+ 0.53 \atop -0.40 }$ & 4.44 $^{+ 0.67 \atop 
      -0.67 }$ & 1450 & 5.81 $^{+ 
      1.14 \atop -0.90 }$ & 3.46 $^{+ 1.17 \atop -0.74 
      }$\\ 
A2199 & 957. & 2.26 $^{+ 0.38 \atop -0.28 }$ 
      & 2.69 $^{+ 0.42 \atop -0.42 }$ & 1621. 
      & 4.98 $^{+ 0.93 \atop -0.67 
      }$ & 4.81 $^{+ 0.37 \atop -0.36 }$\\ 
 A2255 & 1072. & 3.94 $^{+ 0.44 \atop -0.31 }$ & 
      7.13 $^{+ 1.38 \atop -1.37 }$ & 2271. & 
      12.27 $^{+ 1.87 \atop -1.37 }$ 
      & 13.32 $^{+ 1.44 \atop -1.19 }$\\ 
A2589 & 848. & 1.92 $^{+ 0.33 \atop -0.22 }$ & 
      3.03 $^{+ 0.75 \atop -0.75 }$ & 
      1471.& 4.60 $^{+0.84 \atop -0.58 }$ & 
      3.58 $^{+ 3.86 \atop -1.54 }$\\
\hline
\end{tabular}\\[0.5ex]
{\em Model values $M_{tot}(r_{N00})$ and empirical values $M_{N00}$  in $10^{14} \,M_{\astrosun}$, \\
$r_{N00}$  (empirical) in $kpc$ ($N=1,\, 2$)}
\end{table}
\newpage
\clearpage

\begin{table}{\center \bf Table 3. Model values for halo and baryonic masses at  $r_{200}$}\\[0.5ex]
\begin{tabular}{|r||rrrr|cc|cr|} \hline \\[-2.6ex]
{\em Cluster}& $M_{t}$& $M_{sf}$& 
      $M_{sf2}$& $M_{ph1}$& $M_{gas}$& 
      $M_{\ast}$& $f_{\ast}$& $f_{t}$\\ 
			\hline  \hline \\[-2.2ex] 
		   Coma & 11.15 &  6.44 &  1.73 &  4.71 &  1.77 &  0.276 &  0.16 &  6.3\\
 A85 & 8.03 &  4.52 &  1.01 &  3.51 &  1.54 &  0.140 &  
      0.09 &  5.2\\ 
     A400 &  2.27 &  1.36 &  0.45 &  0.91 &  0.26 &  0.084 &  0.32 &   8.7 \\
 IIIZw54 &2.77 &  1.65 &  0.54 &  1.11 &  
      0.25 &  0.080 &  0.32 &  10.9 \\ 
     A1367 &  3.77 &  2.21 &  0.65 &  1.56 &  0.42 &  0.089 &  
      0.21 &  8.9 \\ 
MKW4 & 0.99 &  0.58 &  0.18 &  0.40 &  0.09 &  0.022 &  0.25 &  10.9 \\ 
     ZwCl215 & 8.00 &  4.55 &  1.11 &  3.45 &  1.19 &  0.137 &  0.12 &  6.7 \\ 
     A1650 &  8.15 &  4.69 &  1.24 &  3.45 &  1.10 &  0.162 &  0.15 &  7.4 \\ 
     A1795 &  8.04 &  4.59 &  1.14 &  3.45 &  
      1.15 &  0.14 &  0.12 &  7.0 \\ 
     MKW8 & 2.12 &  1.24 &  0.36 &  0.88 &  
      0.20 &  0.039 &  0.20 &  10.8 \\ 
A2029 & 12.83 &  7.15 &  1.47 &  5.68 &  2.83 &  0.203 &  
      0.07 &  4.5 \\ 
A2052 & 2.58 &  1.50 &  0.43 &  1.07 &  0.31 &  0.059 &  
      0.19 &  8.3 \\ 
			MKW3S &  3.35 &  1.95 &  0.55 &  1.40 &  0.39 &  
      0.072 &  0.18 &  8.5 \\ 
A2065 &10.28 &  5.80 &  1.32 &  4.48 &  1.48 &  
      0.142 &  0.10 &  6.9 \\ 
A2142 & 12.55 &  6.95 &  1.35 &  
      5.60 &  2.59 &  0.159 &  0.06 &  4.8\\
 A2147 &  4.83 &  2.77 &  0.71 &  2.06 &  0.87 &  0.118 &  0.14 &  5.5 \\ 
A2199 & 4.36 &  2.52 &  0.68 &  1.84 &  0.55 &  0.088 &  0.16 &  8.0 \\ 
     A2255 & 10.35 &  5.84 &  1.33 &  4.51 &  1.76 &  0.166 &  0.09 &  5.9 \\ 
     A2589 &  3.99 &  2.33 &  0.68 &  1.65 &  0.52 &  0.105 &  0.20 &  7.7
 \\											
\hline
\end{tabular}\\[0.5ex]
{\em Mass values   in $10^{14} \,M_{\astrosun}$,
 $f_{\ast}=\frac{M_{\ast}}{M_{gas}}, \, f_{t}=\frac{M_{t}}{M_{gas}}(r_{200}), $\\
 $r_{200}$   see tab. 2}
\end{table}

\clearpage

%
\clearpage

\clearpage

\subsection{\small Comparison with TeVeS and NFW halos \label{subsection comparisons}}
It is surprising   that in the WST-3L  approach the total amount of observed dark matter  $M_{dm}$ seems to be explained by the  energy of the scalar field and the phantom halo, $M_{t}=M_{sf}+M_{ph\,1}$,  $M_{dm} \approx M_{t}$. No missing mass  is left.
In usual relativistic MOND approaches this is not the case for clusters, although it is essentially so for galaxies  \citep{Famaey/McGaugh:MOND}. Based on a study of  about 40 galaxy clusters, R Sanders has proposed the hypothesis of a neutrino component ``between a few times $10^{13}$  and $10^{14}\, M_{\astrosun}$'', mostly  concentrated close to the center of the cluster, in a region up to twice the core radius, supplementing the baryonic mass and  the phantom energy of the TeVeS model \citep[p. 902]{Sanders:2003}. In our approach,  this  hypothesis  is unnecessary. Where does this difference arise from?

A model calculation for the Coma cluster, evaluating (\ref{rho-ph}) for the transition function $\mu_2=x(1+x^2)^{-\frac{1}{2}}$ (\ref{mu function}) and the corresponding $\nu_2$  (\ref{nu function}) used by Sanders in \citep{Sanders:2003}, shows that the  neutrino core had to be tuned to  $M_{\nu}\approx \- 1.8\times 10^{14}\, M_{\astrosun}$ in order to give  agreement with the empirical value $M_{dm}(r_{500})=M_{500} - M_{bar}(r_{500}) \approx 4.7 \times 10^{14}\, M_{\astrosun}$, where  in this framework ``$r_{500}$'', ``$r_{200}$'' are to be  defined   by a formal convention with regard to a (fictitious) NFW halo.

Table 3 shows  the amount of scalar field energy up to a radius $r \approx r_{200}$ in the WST model. It varies between about $2\times 10^{13}$ and $9\times 10^{14}\, M_{\astrosun}$,  i.e., roughly in the range found necessary by R. Sanders for the (hypothetical) neutrino halo. Moreover, a comparison of the transition functions $\mu_w(r)$ (\ref{mu-w}) and $\mu_2(x)$ shows that  the gross phantom energy $\rho_{t}$ of the WST approach is larger than in  an  $\mu_2$-MOND model \citep{Scholz:2015MONDlike}. 
Roughly half of the missing mass of Sanders' model is covered by this effect, {\em  the other half is due to the scalar field halo of the system of    galaxies} up to $r_{200}$. 

\begin{figure}
	\centering
		\includegraphics[width=1.00\textwidth]{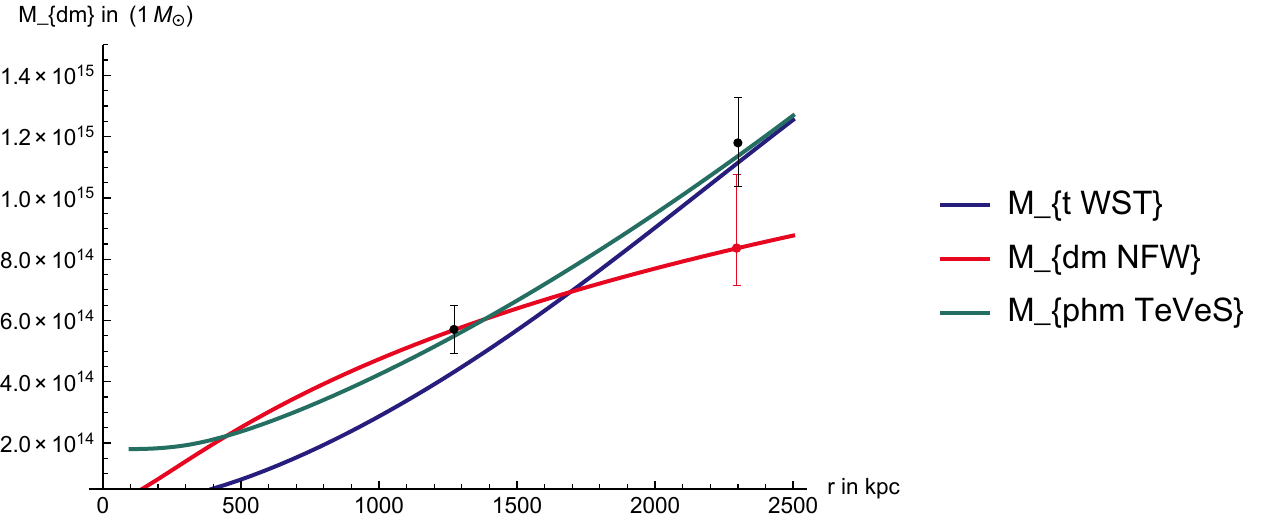}
	\caption{Comparison of dark/transparent/phantom mass halos for  Coma in  NFW, WST and  TeVeS models, \, 
free parameters of	halos for NFW and TeVeS ($\mu_2$ with neutrino core)  adapted to mass data (black error bars) at 
	$r_{500}=1280\, kpc$}
	\label{figure CompNFW}
\end{figure}

A comparison of the two MOND-like approaches is given in  fig. \ref{figure CompNFW}. Here one has to keep in mind that the TeVeS-$\mu_2$ model has a free adaptable parameter (mass of the neutrino core), while the WST has not. The general profiles of the ``dark matter''  halos of both models are  similar. The TeVeS-$\mu_2$  mass starts from a higher  socle because of its neutrino core;  the WST transparent mass starts from a lower initial value but  rises faster because of the increasing contribution of  the scalar field halo of the galaxies.

On the other hand,  the best known  profile for dark matter distribution,  used in most structure formation simulations, is the NFW halo (Navarro/ Frank/White).  Its  profile is 
\beq  \rho (r)= \frac{\rho_o}{\frac{r}{r_c}(1+\frac{r}{r_c})^2 } \; , \label{NFW halo}
\eeq
with density parameter $\rho_o$ and core radius $r_c$ (at which the density has reduced to half the reference value).
 For a first  comparison of the   interior halos we take $r_c\approx 180\, h^{-1}\, kpc $ ($h=0.7$), following
\citep{Geller_ea:1999},\footnote{In studies of the { exterior} halo of the Coma cluster,  Geller e.a. have found  fit values  $0.182^{+0.03 \atop -0.03}, \, 0.167^{+0.029  \atop -0.029}, \, 0.192^{+0.035  \atop -0.035}$ in  units $h^{-1}Mpc$ for $r_c$ \citep{Geller_ea:1999}.} 
and determine the central density parameter $\rho_o$ such that the total integrated mass $M_{NFW}$ assumes the empirical value of \citep{Reiprich/Zhang_ea:2011} at $r_{500}=1289\, kpc$ (fig. \ref{figure CompNFW}). The error intervals for $r_{500}$ give upper and lower model values for $\rho_o$ and corresponding  model error bars (red) for $M_{NFW}(r_{200})$.  This version of the NFW halo  satisfies one set of empirical  data  by construction (at $r_{500}$); our main interest thus goes to the other empirical data set available  at $r_{200} = 2300\, kpc$.  The NFW model error interval overlaps with the empirical error bar at $r_{200}$ and with the error interval of the WST model.

 There is a conspicuous  difference between the  mass profiles of the  NFW halo on one side and the halo profiles of WST or $\mu_2$-TeVeS on the other. More and precise empirical data  of the interior halos of galaxy clusters ought to  be able to  discriminate between the two model classes. An empirical discrimination between the two MOND-like approaches would need more, and more precise, profile data. At the moment the Weyl MONDlike model survives the comparison  fairly well, even though it has no free parameter which would allow to  adapt it to the halo data.

\subsection{\small A side-glance at the bullet cluster  \label{subsection bullet}}

At the end of this section let us shed  a  side-glance at the bullet cluster 1E0657-56. It is often claimed that the latter provides direct evidence in favour of  particle dark matter and  rules out  alternative gravity approaches. Our considerations show that this argument is not  compelling. The energy content of the scalar field halos of the  colliding clusters  endows them with  inertia  of their own. The shock of the colliding gas exerts dynamical forces on the gas masses only, not directly on the scalar field halos. During the encounter the halos will   roughly  follow the inertial trajectories of their respective clusters before collision, and they will continue to do so for a while. It will take  time before a re-adaptation of the mass systems and the respective scalar field halos has taken place. Clearly the MOND-approximation is unable to cover such violent dynamical processes. It  describes only the relatively stable states before collision and -- in some distant future -- after collision. But a separation of halos and gas masses for a (cosmically ``short'') period is to be expected, just like in the case of a particle halo with appropriate clustering properties. 

For the time being, the cluster  1E0657-56 does not help to decide between the overarching alternative  research strategies, particle dark matter or alternative gravity. It may be able to do so, once  the dynamics of gas and of the halos has  been modeled  with sufficient precision in both approaches. Then a proper comparison can be made.  But that is an overtly complicated task. It  seems more likely that other types of empirical evidence will offer a simpler path to a differential evaluation of the two strategies  and  help  clarifying  the  alternative.

\section{ Discussion  \label{section discussion}}
We have analyzed a three-component halo model for clusters of galaxies, consisting of 
\begin{itemize}
	\item[$\;\; (i)$] the scalar field energy induced by the overall baryonic matter in the barycentric rest system of a cluster (under abstraction of the discrete structure of  the star matter clustering in freely falling galaxies) (\ref{scalar field halo baryonic matter}), with integrated mass equivalent $M_{s\hspace{-0.1em}f\, 1 }$,
	\item[$\;(ii)$] an additional contribution to the scalar field energy,  forming around the freely falling galaxies (\ref{scalar field halo galaxies}), integrating to $ M_{s\hspace{-0.1em}f\, 2}$,
	\item[$(iii)$] and the phantom energy of the total baryonic mass in the barycentric rest system of the cluster, due  to the additional acceleration of the Weylian scale connection in Einstein gauge (\ref{phantom halo baryonic matter}) with  mass equivalent $M_{ph\,1}$.
\end{itemize}
The first two components add up to a real energy content of the scalar field with mass equivalent $M_{sf}= M_{s\hspace{-0.1em}f\, 1 }+M_{s\hspace{-0.1em}f\, 2}$, the third one, $M_{ph\,1}$, arises from a theoretical attribution in a Newtonian perspective and has fictitious character. 
The mass equivalent of the integrated energy components  combines to a total dark-matter-like quantity $M_{t}= M_{s\hspace{-0.1em}f\, 1 } +  M_{s\hspace{-0.1em}f\, 2} + M_{ph\,1}$  (\ref{total scalar field halo}).

All three components arise  from gravitational effects of the cluster's  baryonic mass $M_{bar}=M_{gas}+M_{star}$ in the framework of a Weyl geometric scalar tensor theory of gravity (\ref{modified Hilbert term}), with its scale connection  as the specific difference to Riemannian geometry (\ref{varphi}), (\ref{gauge transformation}).\footnote{In Weyl geometric scalar tensor theory the scalar field $\phi$ is the new dynamical variable, while the   {\em scale curvature}  $d\varphi=f$ vanishes. The latter would be an even more striking,  and even irritating,  difference to Riemannian geometry; in integrable Weyl geometry it plays no role. } 
A second speciality of the theoretical framework is the cubic  kinetic term of the Lagrangian (\ref{cubic Lagrangian}),   analogue to the AQUAL approach but in scale covariant form. Observable quantities are directly given by the model in Einstein gauge (\ref{Einstein gauge}).

The total dynamical mass  of the model, $M_{tot}=M_{bar}+M_{t}$ (\ref{total mass}) has been heuristically confronted with the  empirical values for of 17+2 galaxy clusters given in \citep{Lagana/Reiprich_ea:2011}, \citep{Reiprich/Zhang_ea:2011}, \citep{Reiprich/Zhang_ea:Corr}, complemented by data from \citep{Reiprich:Diss}.
The problem of data transfer between different theoretical frameworks (in particular between Einstein gravity /$\Lambda CDM$ and WST-3L)  leads to a certain caveat with regard to an uncorrected taking over of the values for the total mass. But it does not seem to obstruct a meaningful  first comparison of the WST halo model with available mass data of clusters collected in the Einstein/$\Lambda CDM$ framework. 

The result of this comparison shows a surprisingly good agreement of the total mass predicted by the model $M_{tot}$, on the basis of data for the baryonic mass components,  with the observed total mass $M_{500}$  (measured at the main reference distance $r_{500}$). Moreover the model shows an  acceptably good agreement with additional empirical values at the distance $r_{200}$ given in \citep{Reiprich:Diss} (determined on  a slightly less refined data basis and evaluation method). 
For 14 clusters the model predicts values for $M_{tot}(r_{500})$ with error intervals (due to the observational errors for the baryonic data) which  overlap  with the empirical error intervals of $M_{500}$. The Coma cluster is among them.  Three clusters have overlapping error intervals  in the $2 \sigma$ range. The remaining two  are  outliers and have been identified as such already in \citep{Reiprich/Zhang_ea:2011}.

In the result we have found  empirical support for   {\em conjecturing} that the {\em observed dark matter} at galaxy cluster level {\em may be due to the transparent halo of the scalar field and the phantom halo} of   the scale connection of WST (\ref{conjecture}).

Details for the constitution of the total transparent matter halo from its specific components  $(i), \, (ii), \, (iii)$ have been investigated for the Coma cluster (section \ref{subsection Coma}). They seem to be exemplary for the whole collection of galaxy clusters. A particular feature of the model is the {\em scalar field energy formed in the inter-spaces between the galaxies}.  Its integrated energy contribution surpasses the  baryonic mass between 1 and 1.5 $Mpc$ (see fig. \ref{SF star comparative} and table 3, col. 4). It is crucial for this model's capacity to explain the total dynamic mass on purely gravitational grounds, without any additional dark matter component. An overall comparison  with R. Sanders' TeVeS-MOND model for galaxy clusters and the NFW halo is given in section \ref{subsection comparisons}. 

 At the moment the Weyl geometric scalar tensor model with a cubic term in the kinetic Lagrangian of the scalar field  fares  well in all the mentioned respects. 
It would be very helpful if astronomers decided to evaluate old or new raw data in the framework of WST. That could lead to an empirical discrimination of the different models. But already independent of the outcome of such a revision, the scalar field $\phi$ and the scale connection $\varphi_{\mu}$ of WST  have a remarkable property from a theoretical point of view. They complement the classical  Einstein-Riemannian expression for the gravitational structure, the metric field $g_{\mu \nu}$, by a  feature which carries a   proper energy-momentum tensor (\ref{ThetaI}), (\ref{ThetaII}). The {\em energy momentum of the scalar field}   plays a crucial role for the constitution of the transparent matter halo. In the present approach it  seems to express the {\em self-energy of the extended gravitational structure}. It remains to be seen whether this is more than a model artefact.

\section{ Appendix} 
\subsection{\small  Energy component of the scalar field}
In scalar field (Einstein) gauge  $D_{\nu}\phi= \partial_{\nu}\phi - \varphi_{\nu}\phi\doteq -\phi \varphi_{\nu} \doteq\phi \partial_{\nu}\omega \doteq \phi D_{\nu}\omega$,  and  $D_{\nu}\phi^2= 2\phi D_{\nu}\phi\doteq 2 \phi^2 \partial_{\nu}\omega$.
Simliarly
\[ D_{\mu}D^{\nu}\phi^2 \doteq 2 D_{\mu}(\phi^2 \partial^{\nu}\omega) \doteq 2 (D_{\mu}\phi^2 \partial^{\nu} \omega + \phi^2 D_{\mu} \partial^{\nu} \omega)\, .
\]
Moreover, 
\[ D_{\mu}\partial^{\nu} \omega =  \nabla\hspace{-0.2em}_{\mu} \partial^{\nu} \omega   - 2 \varphi_{\mu} \partial^{\nu} \omega   =  _g\hspace{-0.5em}\nabla\hspace{-0.2em}_{\mu} \partial^{\nu} \omega + \delta_{\mu}^{\nu} \varphi_{\lambda} \partial^{\lambda} \omega +  \delta_{\lambda}^{\nu} \varphi_{\mu} \partial^{\lambda} \omega - g_{\mu \lambda } \varphi^{\nu} \partial^{\lambda} \omega + 2 \partial_{\mu} \omega \partial^{\nu} \omega
\]
leads to
\beqa D_{\mu}D^{\nu}\phi^2   &\doteq& 2 \phi^2 (2 \partial_{\mu} \omega \partial^{\nu} \omega + _g\hspace{-0.5em}\nabla\hspace{-0.2em}_{\mu} \partial^{\nu} \omega + 2 \partial_{\mu}\omega \partial^{\nu}\omega - \delta_{\mu}^{\nu}\, \partial_{\lambda}\omega\partial^{\lambda}\omega ) \\
	&\doteq& 2 \phi^2 (  _g\hspace{-0.2em}\nabla\hspace{-0.2em}_{\mu} \partial^{\nu} \omega + 4 \partial_{\mu}\omega \partial^{\nu}\omega - \delta_{\mu}^{\nu}\, \partial_{\lambda}\omega\partial^{\lambda}\omega ) \, .
\eeqa
Thus
\beq  D_{\lambda}D^{\lambda} \phi^2 \doteq  2 \phi^2 \,  _g\hspace{-0.2em}\nabla\hspace{-0.2em}_{\lambda} \partial^{\lambda} \omega 
\eeq
Following   (\ref{ThetaI}), (\ref{ThetaII}),  (\ref{H^2}), the energy momentum of the scalar field, 
\[ \Theta = \theta^{(I)} + \theta^{(II)} = (\xi \phi)^2\,T^{(\phi)}   =  (\xi \phi)^2\,  (T^{(I)} + T^{(II)}) \, ,
\] 
becomes
\beqa  T^{(I)} &\doteq & - (8 \pi G)^{-1} \left(2 \,  _g\hspace{-0.2em}\nabla\hspace{-0.2em}_{\lambda} \partial^{\lambda} \omega   - 3\,\partial_{\lambda}\omega\partial^{\lambda}\omega  -  \frac{2}{3} \tilde{a}_o |\nabla \omega|^3 + \frac{\lambda}{4} H^2  \right)\, g  \\
 T_{\mu \nu}^{(II)} &\doteq& \xi^2 D_{\mu}D_{\nu} \phi^2 - 2 \frac{\partial L_{\phi}}{\partial g^{\mu \nu}} \\
&\doteq&  (8 \pi G)^{-1} \left(  _g\hspace{-0.2em}\nabla\hspace{-0.2em}_{\mu} \partial_{\nu} \omega  + (1-   \tilde{a}_o |\nabla \omega|) \partial_{\mu}\omega  \partial_{\mu}\omega -  \partial_{\lambda}\omega\partial^{\lambda}\omega\,  g_{\mu \nu} \right) \, .
\eeqa
In the static  case the first two terms of $ T_{\mu \nu}^{(II)}$ vanish and the total energy component of the scalar field turns into
\beq T_{oo}^{\phi} \; \doteq \; (4 \pi G)^{-1} \left (  _g\hspace{-0.2em}\nabla\hspace{-0.2em}_{\lambda} \partial^{\lambda} \omega   - \partial_{\lambda}\omega\partial^{\lambda}\omega  -  \frac{1}{3} \tilde{a}_o |\nabla \omega|^3 + \frac{\lambda}{8} H^2 \right)\, .
\eeq 
Assuming conditions under which the terms of  cosmological orders of magnitude and those of order $|\nabla \omega|^2$ can be neglected, we arrive at
\[ T_{oo}^{\phi}\; \dot{\approx} \; (4 \pi G)^{-1} \,  _g\hspace{-0.2em}\nabla\hspace{-0.2em}_{\lambda} \partial^{\lambda} \omega \, ,  
\]
 like in equation (\ref{SF energy density}) of the main text. 
In the central symmetric case  (\ref{solution non-linear Poisson equ}) implies  $\omega (r)= \sqrt{G M \tilde{a}_o} \ln r$.  In spherical coordinates  $\nabla \omega= (0,\frac{\sqrt{GM \tilde{a}_o}}{r}, 0 , 0)$ and 
\beq \nabla^2 \omega =   _g\hspace{-0.2em}\nabla\hspace{-0.2em}_{\lambda} \partial^{\lambda} \omega \doteq \frac{\sqrt{GM \tilde{a}_o}}{r^2}
\eeq
Because of $\partial_{1}\omega \partial_{1}\omega = \frac{GM \tilde{a}_o}{r^2}  \ll \frac{\sqrt{GM \tilde{a}_o}}{r^2} $ the approximation (\ref{SF energy density}) is justified. 

\subsection{\small Remarks on the numerical implementation} 
The calculations in sections 3.2ff. have been implemented in {\em Mathematica 10} and run on a PC. Integrations of the mass values have been realized by numerical interpolation routines in distance intervals of $100\, kpc$. A comparison with refined distance intervals $10\, kpc$ showed differences at the order of magnitude $10^{-4}$ of the respective values, thus below the rounding precision. 

The fading out  for the scalar field and phantom halos beyond $r_{200}$ has been modeled by the cubic expression:
\beq  f(x, A,B) = \chi (x; -\infty,A) + \left(\frac{1}{1+\frac{x-A}{B}}\right)^3\chi (x; A,\infty) \, , \label{fading out}
\eeq
with $\chi (x;a,b)$ the characteristic function of the interval $[a,b]$. The fading out of $f(x,A,B)$ starts at $A$ and  declines to $\frac{1}{2}$ at $A+B$. In our case we start the fading out close to the virial radius, $A=1.1\, r_{200}$ and set $B=0.5\, r_{200}$.
%
\clearpage

\small
\addcontentsline{toc}{section}{\protect\numberline{}Bibliography}
 \bibliographystyle{apsr}
  \bibliography{a_lit_mathsci,a_lit_hist}

\end{document}